\documentclass[fleqn, usenatbib]{mnras}

\usepackage{newtxtext, newtxmath}
\usepackage[T1]{fontenc}
\usepackage{ae, aecompl}
\usepackage{graphicx}
\graphicspath{{Figures/}}
\usepackage{tabularx}
\usepackage{amsmath}
\usepackage{amssymb}
\usepackage{gensymb}
\usepackage{wasysym}
\usepackage{color}
\usepackage{multirow}
\usepackage{xcolor}

\newcolumntype{Y}[1]{>{\centering\arraybackslash}p{#1}}
\newcolumntype{Z}[1]{>{\raggedright\arraybackslash}p{#1}}
\newcommand{\cshift}{\ensuremath{\langle w \rangle}}

\title[Relaxation with \textsc{Mock-X}]{Studying galaxy cluster morphological metrics with \textsc{Mock-X}}

\author[K. Cao et al.]{
Kaili Cao \href{https://orcid.org/0000-0002-1699-6944}{\includegraphics[scale=0.1]{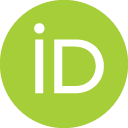}},$^{1}$\thanks{E-mail: \href{mailto:kailicao@mit.edu}{kailicao@mit.edu}}
David J. Barnes \href{https://orcid.org/0000-0002-6583-9478}{\includegraphics[scale=0.1]{ORCIDiD.png}},$^{1}$\thanks{E-mail: \href{mailto:djbarnes@mit.edu}{djbarnes@mit.edu}}
Mark Vogelsberger \href{https://orcid.org/0000-0001-8593-7692}{\includegraphics[scale=0.1]{ORCIDiD.png}},$^{1}$
\\
$^1${Department of Physics, Kavli Institute for Astrophysics and Space Research, Massachusetts Institute of Technology, Cambridge, MA 02139, USA}\\
}

\date{Accepted XXX. Received YYY; in original form ZZZ}

\pubyear{2020}

\begin{document}
\label{firstpage}
\pagerange{\pageref{firstpage}--\pageref{lastpage}}
\maketitle

\begin{abstract}
Dynamically relaxed galaxy clusters have long played a role in galaxy cluster studies because it is thought their properties can be reconstructed more precisely and with less systematics.
As relaxed clusters are desirable, there exist a plethora of criteria for classifying a galaxy cluster as relaxed.
In this work, we examine $9$ commonly used observational and theoretical morphological metrics extracted from $54,000$ \textsc{Mock-X} synthetic X-ray images of galaxy clusters taken from the IllustrisTNG, BAHAMAS and MACSIS simulation suites.
We find that the simulated criteria distributions are in reasonable agreement with the observed distributions.
Many criteria distributions evolve as a function of redshift, cluster mass, numerical resolution and subgrid physics, limiting the effectiveness of a single relaxation threshold value.
All criteria are positively correlated with each other, however, the strength of the correlation is sensitive to redshift, mass and numerical choices.
Driven by the intrinsic scatter inherent to all morphological metrics and the arbitrary nature of relaxation threshold values, we find the consistency of relaxed subsets defined by the different metrics to be relatively poor.
Therefore, the use of relaxed cluster subsets introduces significant selection effects that are non-trivial to resolve.
\end{abstract}

\section{Introduction}
\label{sec:intro}
Galaxy clusters are the most massive collapsed structures observed in the Universe at the current epoch.
The product of runaway gravity acting on cosmic timescales, clusters consist of dark matter, $10^8\, \mathrm{K}$ hot gas and thousands of galaxies \citep[e.g.][]{White1993,Evrard1997,Kravtsov2005}.
Galaxy clusters originate from the high-amplitude tail of the primordial fluctuations present in the early Universe.
As such, the number of clusters as a function of mass and redshift is highly sensitive to the fundamental cosmological parameters that govern the Universe \citep[e.g.][]{Allen2011,Kravtsov2012,Pratt2019}.
Their hierarchical formation over cosmic time also enables them to place stringent constraints on the nature of dark energy \citep[e.g.][]{Weinberg2013}.
The estimates placed by galaxy clusters on fundamental cosmological parameters are complementary and often orthogonal to other probes of the early Universe \citep[e.g.][]{Mantz2014,deHaan2016,Bocquet2019}.

The next generation of observational facilities is currently either undertaking or preparing surveys that will be transformational for cluster science.
Facilities like SPT-3G \citep{Benson2014}, eROSITA \citep{Merloni2012}, Euclid \citep{Laureijs2011}, LSST \citep{LSST2009}, and the Simons observatory \citep{SO2019} will yield large samples of clusters at sub-millimeter, infrared, optical, and X-ray wavelengths.
Expected to yield samples of up to $100,000$ clusters, these surveys will increase the number of known clusters by two orders of magnitude \citep[e.g.][]{Borm2014,Clerc2018,Mantz2019}.
Combined with comprehensive followup campaigns, these surveys will yield a meticulously detailed, multiwavelength picture of cluster formation over cosmic time.

The fundamental requirement for clusters to constrain conditions in the early Universe are robust, low-scatter mass estimates \citep[e.g.][]{Mantz2019,Pratt2019}.
For the majority of clusters, these masses will be relatively derived using mass-observable scaling relations.
However, to calibrate and self-consistently derive these scaling relations a small fraction of the clusters will require highly detailed observations to directly estimate the mass.
For these absolute mass measurements, dynamically relaxed clusters have often play a special role \citep[e.g.][]{Allen2001,Vikhlinin2006,Arnaud2007,Rapetti2008,Vikhlinin2009a,Mantz2010a,Mantz2016}.
Clusters are dynamical active objects, with merging substructures and in-falling material continually driving bulk and turbulence motions within their internal structure. 
It is believed that only in the most regular galaxy clusters can the large-scale $3$D properties be reliably recovered from the projected $2$D observables.
The masses of relaxed clusters are thought to be recovered with higher precision, less bias and smaller systematic uncertainties.

Traditionally, the dynamical state of a galaxy cluster has been determined via visual examination \citep[e.g.][]{Baier1996,Jones1999}. 
A regular morphology and the presence of strong central emission, associated with a cool-core, being the key requirements for a cluster to be classified as relaxed.
However, this approach has an inherent lack of objectivity and it does not scale to the size of future samples.
The alternative approach is to compute morphological indicators from cluster images \citep[e.g.][]{Mohr1995,Buote1995,Poole2006, Santos2008,Maughan2008,Okabe2010,Nurgaliev2013,Mantz2015,Lovisari2017}.
These methods yield objective and reproducible processes for classifying clusters as relaxed.
The drawback of these techniques is that they fundamentally reduced to a thresholding exercise, with the threshold values set by the study.
The choice of classification feature used depends strongly on the quality of the data, with redshift, signal-to-noise, points source masking and unexposed focal plane issues all complicating the task.
There have been previous studies exploring the performance of different features \citep[e.g.][]{Rasia2013}, but little work to assess the consistency of relaxed subsets yielded by different criteria.

Theoretically, numerical simulations should provide clarity as the cluster properties are known exactly.
With ever-increasing computational power and the continual development of calibrated subgrid physical models, hydrodynamical simulations are now capable of simulating large cosmological volumes that are increasingly realistic \citep{LeBrun2014,Vogelsberger2014, Schaye2015,Dubois2016,McCarthy2017,Springel2018, Dave2019,Vogelsberger2020}, i.e. the properties of collapsed haloes broadly match the observations.
Through these simulations and dedicated zoom simulation campaigns \citep[e.g.][]{Planelles2013,Rasia2015,Barnes2017a,Barnes2017b,Cui2018,Henden2018} there are increasingly large, realistic simulated galaxy cluster samples.
However, theoretical relaxation criteria use a different set of metrics measured within $3$D volumes \citep[e.g.][]{Neto2007, Duffy2008, Klypin2011, Dutton2014, Klypin2016, Barnes2017b}, rather than projected apertures.
Additionally, theoretical approaches aim to exclude the most disturbed objects, rather than selecting a subset of the most relaxed clusters.
These methods also reduce to a thresholding exercise, with a combination of criteria typically used in numerical studies.
Therefore, it is unclear if theoretical subsets of relaxed clusters are consistent with observational subsets.

The goal of this paper is to examine various observational and theoretical morphological criteria commonly used to classify clusters as relaxed.
Using the \textsc{Mock-X} analysis framework \citep{Barnes2020}, we measure the relaxation criteria of cluster samples selected from the BAHAMAS \citep{McCarthy2017}, MACSIS \citep{Barnes2017a} and IllustrisTNG \citep{Nelson2018, Pillepich2018b, Springel2018, Naiman2018, Marinacci2018} simulations via synthetic X-ray images \citep{Barnes2020} at a range of redshifts.
We compare the simulated and observed criteria distributions before examining how these criteria evolve as a function of mass, redshift, subgrid physics and numerical resolution.
We then explore the correlation between criteria and study the consistency of relaxed cluster subsets selected with different morphological metrics.

The rest of this paper is structured as follows.
In Section \ref{sec:methods} we present our numerical method, including a brief overview of the hydrodynamical simulations, the \textsc{Mock-X} synthetic image framework, the relaxation criteria studied in this work, and our statistical methods.
The results regarding the distribution, correlation and consistency of the relaxation parameters are presented and discussed in Section \ref{sec:dist}, Section \ref{sec:corr} and Section \ref{sec:consis}, respectively.
Finally, our conclusions are summarized in Section \ref{sec:concs}.

\section{Methods}
\label{sec:methods}
Throughout this work we utilize the IllustrisTNG $(302)^{3}\,\mathrm{Mpc}^{3}$ volume \citep{Marinacci2018,Naiman2018,Nelson2018,Pillepich2018b,Springel2018}, the reference \textit{Planck} cosmology run of the BAHAMAS simulation suite \citep{McCarthy2017} and the MACSIS zoom simulation suite \citep{Barnes2017a}.
The cosmological parameters adopted for these runs are summarized in Table \ref{tab:cosmo}.
BAHAMAS and MACSIS both assume a \citet{Planck2014} cosmology, with the small differences between them depending on whether the \textit{Planck} data is combined with BAO, \textit{WMAP} polarization and high multipole moments experiments.
IllustrisTNG assumes a \citet{PlanckXIII2016} cosmological model.
The minor differences in adopted cosmology between the simulations used in this work have a negligible impact on the results presented in this paper.
We now briefly outline the subgrid physical models used in the three simulation sets, our cluster selection criteria, the MOCK-X synthetic image framework for generating X-ray images, the chosen morphological criteria and how they were computed, and, finally, the statistical methods used in this work.
We refer the interested reader to the relevant papers in each subsection for further details.

\renewcommand\arraystretch{1.2}
\begin{table}
 \caption{Table summarizing the $\Lambda$CDM model parameters adopted by the hydrodynamical simulation used in this work.}
 \centering
 \begingroup
 \setlength{\tabcolsep}{4pt}
 \begin{tabular}{c c c c c c c}
  \hline
  Simulation & $\Omega_\mathrm{m}$ & $\Omega_\mathrm{b}$ & $\Omega_{\Lambda}$ & $\sigma_{8}$ & $n_\mathrm{s}$ & $h$ \\
  \hline
  BAHAMAS & $0.3175$ & $0.0490$ & $0.6825$ & $0.834$ & $0.9624$ & $0.6711$ \\
  MACSIS & $0.307$ & $0.04825$ & $0.693$ & $0.8288$ & $0.9611$ & $0.6777$ \\
  IllustrisTNG & $0.3089$ & $0.0486$ & $0.6911$ & $0.8159$ & $0.9667$ & $0.6774$ \\
  \hline
 \end{tabular}
 \endgroup
 \label{tab:cosmo}
\end{table}
\renewcommand\arraystretch{1.0}

\subsection{Cosmological hydrodynamical simulations}
\label{sec:sim_projs}
In this work, we select cluster samples from all three resolution levels of the TNG300 simulation, the reference BAHAMAS volume, and all MACSIS simulations.
The key numerical parameters of these simulations are summarized in Table \ref{tab:sim_params_num}.
The different calibrated subgrid physics models used for IllustrisTNG and BAHAMAS/MACSIS enables a study of the impact of the chosen subgrid method on morphological criteria.
Therefore, we now briefly outline the subgrid models.

\subsubsection{IllustrisTNG}
\label{sec:IllustrisTNG}
The IllustrisTNG project \citep{Nelson2018,Pillepich2018b,Springel2018,Naiman2018,Marinacci2018}
is a follow-up project to the Illustris simulation \citep{Genel2014,Vogelsberger2014,VogelsNat2014,Sijacki2015}, and is composed of $50^{3}$, $100^{3}$ and $300^{3}\,\mathrm{Mpc}^{3}$ periodic volumes run with an updated galaxy formation model \citep{Pillepich2018a}.
It evolves the magnetohydrodynamic equations using the moving-mesh code \textsc{Arepo} \citep{Springel2010}.
It has an extended chemical evolution scheme and a re-calibrated SN wind model \citep{Pillepich2018a}.
The feedback model has been redesigned and includes a new radio mode active galactic nuclei (AGN) feedback scheme \citep{Weinberger2017}.
Further refinements to the numerical scheme that improve its convergence properties are also included \citep{Pakmor2016}.
In this work, we exclusively use the $300^{3}\,\mathrm{Mpc}^{3}$ volume, making use of all three resolution levels.
The mass (spatial) resolution decreases by a factor $8$ ($2$) from level $1$ to $2$ and from $2$ to $3$, which enables a study of the impact of numerical resolution on the morphological parameters.
We note that the IllustrisTNG model is only calibrated for the highest resolution simulation.
At cluster scales, the IllustrisTNG model has been shown to reproduce a realistic intracluster medium (ICM), with low-redshift cool-core metrics in reasonable agreement with observed low-redshift clusters \citep{Barnes2018,Barnes2019}.

\renewcommand\arraystretch{1.1}
\begin{table}
 \caption{Table of numerical parameters for the simulations used in this work. $L_\mathrm{box}$ is the side length of the cubic simulation boxes; $N_\mathrm{DM}$ ($N_\mathrm{GAS}$) is the number of dark matter (gas) particles; $m_\mathrm{DM}$ ($m_\mathrm{baryon}$) is the (initial) mass of a dark matter (gas) particle; $\epsilon_\mathrm{DM, stars}$ ($\epsilon_\mathrm{gas, min}$) is the minimum Plummer equivalent gravitational softening length for the collisionless (gas) particles. The gravitational softening length is fixed to the minimum value in physical coordinates below $z = 3$ ($z = 1$) and fixed in comoving coordinates at higher redshifts in BAHAMAS and MACSIS (IllustrisTNG).}
 \centering
 \begingroup
 \setlength{\tabcolsep}{1.1pt}
 \begin{tabular}{c c c c c c c c}
  \hline
  Simulation & $L_\mathrm{box}$ & $N_\mathrm{DM}$ & $N_\mathrm{GAS}$ & $m_\mathrm{DM}$ & $m_\mathrm{baryon}$ & $\epsilon_\mathrm{DM}$ & $\epsilon_\mathrm{gas}$ \\
  & [Mpc$/h$] & & & [$\mathrm{M}_{\astrosun}/h$] & [$\mathrm{M}_{\astrosun}/h$] & [kpc$/h$] & [kpc$/h$] \\
  \hline
  BAHAMAS & $400$ & $1024^3$ & $1024^3$ & $4.45 \!\times\! 10^9$ & $8.12 \!\times\! 10^8$ & $4.0$ & $4.0$ \\
  MACSIS & --- & --- & --- & $4.4 \!\times\! 10^9$ & $8.0 \!\times\! 10^8$ & $4.0$ & $4.0$ \\
  TNG300-L1 & $205$ & $2500^3$ & $2500^3$ & $4.0 \!\times\! 10^7$ & $7.6 \!\times\! 10^6$ & $1.0$ & $0.25$ \\
  TNG300-L2 & $205$ & $1250^3$ & $1250^3$ & $3.2 \!\times\! 10^8$ & $5.9 \!\times\! 10^7$ & $2.0$ & $0.5$ \\
  TNG300-L3 & $205$ & $625^3$ & $625^3$ & $2.5 \!\times\! 10^9$ & $4.8 \!\times\! 10^8$ & $4.0$ & $1.0$ \\
  \hline
 \end{tabular}
 \endgroup
 \label{tab:sim_params_num}
\end{table}
\renewcommand\arraystretch{1.0}

\subsubsection{BAHAMAS}
\label{sec:BAHAMAS}
The BAHAMAS project \citep{McCarthy2017} was devised to study large-scale structure (LSS) cosmology with self-consistent hydrodynamical simulations.
Built upon the success of OWLS \citep{Schaye2010} and Cosmo-OWLS \citep{LeBrun2014}, BAHAMAS evolves the hydrodynamic equations using traditional smooth particle hydrodynamics (SPH) via the Lagrangian TreePM-SPH code \textsc{gadget3} \citep[last described in][]{Springel2005}.
The subgrid galaxy formation model includes radiative cooling via a cloudy lookup table \citep{WiersmaSchayeSmith2009} and stochastic star formation that by construction reproduces the Kennicutt-Schmidt law \citep{SchayeDallaVecchia2008}.
Stellar evolution and chemical enrichment are computed via the prescription of \citet{Wiersma2009}, and galactic outflows are generated via the kinetic SN feedback model of \citet{DallaVecchiaSchaye2008}.
Supermassive black hole (SMBH) seeding, growth and AGN feedback are calculated using the recipe of \citet{BoothSchaye2009}, a modified version of the techniques developed by \citet{SpringelDiMatteoHernquist2005}.
The reference BAHAMAS \textit{Planck} volume is a $(596)^{3}\,\mathrm{Mpc}^{3}$ volume with an initial gas (dark matter) mass of $1.21\times10^{9}\,\mathrm{M}_{\astrosun}$ ($6.63\times10^{9}\,\mathrm{M}_{\astrosun}$).
The minimum smoothing length of the SPH kernel is set to a tenth of the gravitational softening, which is set to $5.96$ comoving (physical) $\mathrm{kpc}$ for $z>3$ ($z\leq3$).

\subsubsection{MACSIS}
\label{sec:MACSIS}
The MACSIS project \citep{Barnes2017a} is a suite of 390 zoom simulations that target the rarest, most massive clusters expected to form in a $\Lambda$CDM cosmology.
The clusters were selected from a $(3.2)^{3}\,\mathrm{Gpc}^{3}$ cubic periodic parent simulation \citep[see][for more details]{Barnes2017a}.
The clusters were then resimulated at a higher resolution using the zoom simulation technique \citep{KatzWhite1993,Tormen1997}, ensuring the high-resolution region is uncontaminated to at least $5\,r_{\mathrm{500,crit}}$\footnote{The radius $r_{\mathrm{500,crit}}$ denotes the radius of a sphere that encloses a mass $M_{\mathrm{500,crit}}$ and has a mean density equal to 500 times the critical density of the Universe.}.
The mass and spatial resolution of the resimulations were chosen to be an exact match to the BAHAMAS simulation, and the BAHAMAS galaxy formation model was used for the hydrodynamical simulations.
The combination of MACSIS and BAHAMAS enables the morphological parameters to be studied over the complete cluster mass range and differences between the samples are likely driven by the difference in average mass.

\subsubsection{Cluster sample selection}
All simulations used in this work identify haloes via a \textit{Friends-of-Friends} (FoF) percolation algorithm run on the dark matter particles.
A linking length in units of the mean interparticle separation of $b=0.2$ was used.
Baryonic particles are then attached to haloes by locating their nearest dark matter particle.
Bound substructures are then identified via the \textsc{Subfind} algorithm \citep{Springel2001,Dolag2009}.
The most massive bound structure in each FoF group is labelled as a central, with all other bound structures labelled as substructures.
The cluster centre is always defined by the particle with the lowest gravitational potential that is bound to the central object.
For all simulations, we select all clusters with a mass $M_{\mathrm{200,crit}}>10^{14}\,\mathrm{M}_{\astrosun}$ from the snapshots closest to $z=0.1$, $0.3$, $0.5$ and $1.0$.
The minor differences in redshift between the simulations have a negligible impact on the results presented in this paper.
Our selection yields combined samples $3126$, $2760$, $2125$ and $1005$ clusters at $z=0.1$, $0.3$, $0.5$ and $1.0$, respectively.
Table \ref{tab:cnums} summarizes the number of clusters selected as a function of simulation and redshift.
Every selected cluster was then run through the \textsc{Mock-X} framework to generate $6$ projections for every cluster.

\renewcommand\arraystretch{1.25}
\begin{table}
 \caption{Table summarizing the number of clusters selected at each redshift for the different simulation samples.}
 \centering
 \begin{tabularx}{\columnwidth}{Z{1.7cm} Y{1.25cm} Y{1.25cm} Y{1.25cm} Y{1.25cm}}
 \hline
 Simulation & $z=0.1$ & $z=0.3$ & $z=0.5$ & $z=1.0$ \\
 \hline
 TNG300-L1 & 250 & 196 & 149 & 50 \\
 TNG300-L2 & 250 & 191 & 148 & 49 \\
 TNG300-L3 & 242 & 202 & 146 & 46 \\
 BAHAMAS & 1994 & 1781 & 1292 & 482 \\
 MACSIS & 390 & 390 & 390 & 378 \\
 \hline
 \end{tabularx}
 \label{tab:cnums}
\end{table}
\renewcommand\arraystretch{1.0}

\subsection{Synthetic X-ray images}
\label{sec:mockx}
For every cluster in the $5$ samples, we generate synthetic X-ray images using the \textsc{Mock-X} analysis framework.
Three projections are created along the $x$, $y$ and $z$ directions.
A further three projections are produced along the principal axes $A$, $B$ and $C$, defined by the eigenvectors of the inertial tensor
\begin{equation}
 \mathcal{I}_{ij}=\sum_{k=1}^{N_{200}}m_{k}r_{k,i}r_{k,j}\:,
\end{equation}
where $m_{k}$ is the mass of the $k$th cell/particle, $r_{k,i}$ is the $i$th component of the position vector $r_{k}$ in cluster centric
coordinates and the sum is over the number of particles, $N_{200}$, within $r_{\mathrm{200,crit}}$.
By convention, the eigenvalues are arranged such that $A>B>C$.
Each projection is treated as an independent cluster realization throughout this work.

Synthetic X-ray images are created by computing an X-ray spectrum for gas cell/particle within the FoF group using a table of spectral templates.
The table is precomputed using the Astrophysical Plasma Emission Code \citep[APEC;][]{Smith2001} via the \textsc{pyatomdb} module with atomic data from \textsc{atomdb} v3.0.9 \citep[last described in][]{Foster2012}.
The energy range and resolution were set to match the \textit{Chandra} ACIS-I instrument with an energy range of $0.5-10.0\,\mathrm{keV}$
and energy resolution of $150\,\mathrm{eV}$.
The spectra are convolved with the ACIS-I response matrix and the effective area for the desired energy bins is taken from the ancillary response file.
Galactic absorption is modelled via a WABS model \citep{Morrison1983} and we assume a fixed column density of $n_{\mathrm{H}}=2\times10^{20}\,\mathrm{cm}^{-2}$.
Cells/particles whose temperature is $<10^{6}\,\mathrm{K}$, star formation rate is non-zero (i.e. it is following an enforced equation of state), or net cooling rate is positive (i.e. it is increasing in temperature) are discarded from the image-making process because either we do not expect them to significantly emit X-rays or their hydrodynamic properties are unreliable due to the galaxy formation model.

The spectra are then projected down the relevant axis and smoothed onto a square grid with a physical side length of $3\,r_{\mathrm{500,crit}}$.
Observational issues such as chip gaps, the requirement of stitching multiple pointings together, and instrument response variation across the focal plane are neglected.
A \textit{Chandra}-\textit{like} resolution of $0.5\,\mathrm{arcsec}$ is chosen for the pixel resolution.
In this work, we want to assess the fundamental evolution of the morphological criteria with mass and redshift.
Therefore, we assume perfect signal-to-noise and leave the assessment of the impact of noise to future work.

\subsection{Morphological metrics for relaxed clusters}
\label{sec:rlx_crit}
In this section, we introduce the nine relaxation criteria explored in this work.
We outline the theoretical criteria in Section \ref{sec:rlx_params_sim} and the observational metrics in Section \ref{sec:rlx_params_obs}.
The threshold for each morphological metric is taken from the literature reference provided in each section.

\subsubsection{Theoretical measurements}
\label{sec:rlx_params_sim}
Theoretically, there are many ways to define a relaxed halo \citep[see][]{Neto2007, Duffy2008, Klypin2011, Dutton2014, Klypin2016, Barnes2017b}.
In this paper, we study three measurements defined as follows:

\begin{itemize}
\item[(i)] \textit{Centre of mass offset}: A normalized measure of the absolute offset between a cluster's centre of mass, $\mathbf{r}_{\mathrm{com}}$, and its centre of potential, $\mathbf{r}_{\mathrm{pot}}$
\begin{equation}
  X_{\mathrm{off}} = |\mathbf{r}_{\mathrm{pot}}-\mathbf{r}_{\mathrm{com}}|\,/\,r_{\mathrm{500,crit}}\:.
\end{equation}
We compute this criteria for all gas, dark matter and star cells/particles that fall within a $3$D aperture of radius $r_{\mathrm{500,crit}}$.
Those clusters with $X_{\mathrm{off}} < 0.07$ are classified as relaxed \citep{Neto2007}.\\

\item[(ii)] \textit{Substructure mass fraction}: The fraction of mass residing in bound substructures within a $3$D aperture of radius  $r_{\mathrm{500,crit}}$
\begin{equation}
 f_{\mathrm{sub}}=\sum_{i=1}^{N_{\mathrm{sub}}}M_{\mathrm{sub},i}\,/\,M_{\mathrm{500,crit}}\:, 
\end{equation}
where the sum runs over the number of substructures, $N_{\mathrm{sub}}$ and $M_{\mathrm{sub,i}}$ is the mass of the $i$th substructure.
We note that zeroth subhalo is defined as the central object and excluded.
Clusters are classified as relaxed if $f_{\mathrm{sub}} < 0.1$ \citep{Neto2007}.\\

\item[(iii)] \textit{Energy ratio}: The ratio of the kinetic energy, $E_{\mathrm{kin,500}}$, to thermal energy, $E_{\mathrm{thm,500}}$, for all gas cells/particles with a $3$D aperture of radius $r_{\mathrm{500,crit}}$. 
When computing the kinetic energy, the bulk motion of the cluster is removed.
The ratio is defined as
\begin{equation}
 E_{\mathrm{rat}}=E_{\mathrm{kin,500}}\,/\,E_{\mathrm{thm,500}}\:,
\end{equation}
and clusters are defined as relaxed if $E_{\mathrm{rat}} < 0.1$ following \citet{Barnes2017b}.
\end{itemize}

By definition, all these theoretical criteria are always positive and a larger value is more disturbed.
We refer to this type of relaxation parameters as ``negative'' parameters, as their value is negatively correlated to the degree of relaxation.

\subsubsection{X-ray morphological indicators}
\label{sec:rlx_params_obs}
X-ray observations of galaxy clusters provide detailed information on the dynamical state of the ICM.
This had led to the creation of many parameters that measure the morphology of a cluster from its X-ray emission \citep[e.g.][]{Mohr1995, Buote1995, Poole2006, Santos2008, Maughan2008, Okabe2010, Nurgaliev2013, Mantz2015, Lovisari2017}.
For every synthetic image, we compute $6$ commonly used morphological criteria:

\begin{itemize}
\item[(iv)] \textit{Centroid shift}: The standard deviation of the distance between the X-ray peak and the centroid of the X-ray emission for a series of increasingly smaller apertures.
We measure the centroids in $2$D apertures from synthetic X-ray images with radii in the range $0.15-1.0\,r_\mathrm{500,crit}$ and normalize the standard deviation by $r_\mathrm{500,crit}$
\begin{equation}
 \cshift = \frac{1}{r_{\mathrm{500, sim}}}\sqrt{\frac{\sum(\Delta_{i}-\langle\Delta\rangle)^{2}}{M-1}}\:,
\end{equation}
where $M$ is the total number of apertures considered, $\Delta$ is the separation of the centroids, and the angle brackets denote the average.
We note that we use the $r_{\mathrm{500,crit}}$ value derived by the \textsc{Subfind} algorithm and leave the exploration of the impact of mass and radius estimates to future work.
Following \citet{Maughan2012}, cluster is classified as relaxed if $\cshift < 0.006$.\\

\item[(v)] \textit{Power ratios}: The power ratios are the multipole decomposition of the X-ray surface brightness, $S_{\mathrm{X}}$, within a given aperture \citep{Buote1995}.
The third-order multipole, $P_{3}$, provides information about the bimodal nature of the emission and is the most suitable statistic for detecting asymmetries associated with the presence of substructures.
Following previous studies \citep[e.g.][]{Jeltema2005,Jeltema2008,Cassano2010,Weißmann2013,Rasia2013}, we select an aperture $R_{\mathrm{ap}}=r_{\mathrm{500,crit}}$.
The zeroth power ratio is give by
\begin{equation}
 P_{0}=\left[a_{0}\ln(R_{\mathrm{ap}})\right]^{2}\:,
\end{equation}
where $a_{0}$ is the total intensity within the selected aperture.
Any order higher than $m=0$ is defined as
\begin{equation}
P_m = \frac{1}{2m^2R_{\mathrm{ap}}^{2m}}\left(a_m^2+b_m^2\right)\:,
\end{equation}
where the moments $a_{m}$ and $b_{m}$ are given by
\begin{equation}
 a_{m}(r)=\int_{R^{\prime}\leq R_{\mathrm{ap}}}S_{\mathrm{X}}(\mathbf{x^{\prime}})\,R^{\prime}\cos(m\phi^{\prime})\,d^{2}\mathbf{x^{\prime}}\:,
\end{equation}
and
\begin{equation}
 b_{m}(r)=\int_{R^{\prime}\leq R_{\mathrm{ap}}}S_{\mathrm{X}}(\mathbf{x^{\prime}})\,R^{\prime}\sin(m\phi^{\prime})\,d^{2}\mathbf{x^{\prime}}\:,
\end{equation}
respectively, where $\mathbf{x^{\prime}}\equiv(R^{\prime},\phi^{\prime})$ represents the conventional polar coordinates.
Clusters are classified as relaxed via the third-order power ratio if $P_{3}\,/\,P_{0}<10^{-8}$ \citep{Rasia2013}.\\

\item[(vi)] \textit{Photon asymmetry}: The photon asymmetry statistic, $A_{\mathrm{phot}}$, is sensitive to spatial irregularities in a cluster's X-ray emission.
It compares the azimuthal cumulative photon count distribution to a uniform distribution to quantify the extent of its asymmetry.
Fundamentally, it computes the probability that these two distributions are different for a set of predefined annuli using the non-parametric Watson test \citep{Nurgaliev2013}.
Less sensitive to data quality than other measures, it has been used to classify cluster samples that extend to high $(z>0.8)$ redshift \citep{Nurgaliev2017,McDonald2017}.
We compute the photon asymmetry in four annuli bound in the range $[\,0.05,\,0.12,\,0.2,\,0.3,\,1.0\,]\,r_{\mathrm{500,crit}}$.
The weighted average over the four annuli is computed as
\begin{equation}
 A_{\mathrm{phot}}=100\sum_{k=1}^{4}C_{k}\hat{d}_{N_k, C_k}\,/\,\sum_{k=1}^{4}C_{k}\:,
\end{equation}
where $k$ is the current annulus, $C_{k}$ is the number of counts originating from the cluster within the annulus, $N_{k}$ is total number of counts observed in the annulus, and we highlight that in the absence of noise $C_{k}\equiv N_{k}$.
The background-corrected distance estimate, $\hat{d}_{N_k, C_k}$, between the observed distribution and the uniform distribution is given by
\begin{equation}
 \hat{d}_{N_k, C_k}=\frac{N}{C^{2}}\left(U^{2}_{\mathrm{N}}-\frac{1}{12}\right)\:,
\end{equation}
where $U_{\mathrm{N}}$ is the minimum value of Watson's statistic \citep{Waston1961} between the angular cumulative distribution functions integrated over all possible starting angles.
Cluster as classified as relaxed by this statistic if $A_{\mathrm{phot}}<0.15$ \citep{Nurgaliev2013}.\\

\item[(vii)] \textit{Surface brightness peakiness}: Forming part of the symmetry-peakiness-alignment (SPA) joint morphological criteria derived by \citet{Mantz2015}, the surface brightness peakiness is a measure of the central concentration of a cluster's X-ray emission.
To compute this metric, the initial step for a given cluster is to compute a surface brightness scaling motivated by self-similar scaling arguments \citep{Kaiser1986}
\begin{equation}
 f_{\mathrm{S}}=K(z,T,N_{\mathrm{H}})\frac{E^{3}(z)}{(1+z)^{4}}\left(\frac{k_{\mathrm{B}}T}{\mathrm{keV}}\right)\:,
\end{equation}
where $K(z,T,N_{\mathrm{H}})$ is the redshift and temperature corrected bolometric flux in the observed band, $z$ is redshift, $T$ is temperature, $N_{\mathrm{H}}$ is the hydrogen column density, $E(z)\equiv\sqrt{\Omega_{\mathrm{M}}(1+z)^{3}+\Omega_{\Lambda}}$ and $k_{\mathrm{B}}$ is the Boltzmann constant.
This allows a characteristic set of surface brightness levels to be defined in normalized flux units
\begin{equation}
 S_{\mathrm{j}}=0.002\times10^{0.28j}f_{\mathrm{S}}\:,
\end{equation}
where $j=0,1,\ldots,5$.
The surface brightness amplitude and the number of levels were chosen empirically by \citet{Mantz2015} and we have not explored this choice in this work.
Given these characteristic surface brightness levels, the peakiness statistic is defined as
\begin{equation}
 p=\log_{10}\left[(1+z)\frac{\overline{S_{\mathrm{X}}}(\theta\leq\theta_{5})}{f_{\mathrm{S}}}\right]\:,
\end{equation}
where $\overline{S_{\mathrm{X}}}$ is the area-weighted average surface brightness within the isophote $S_{5}$.
The redshift dependence was added empirically, based on previous literature studies, and is something that we will examine in this work.
We adopt the same threshold as \citet{Mantz2015}, classifying clusters as relaxed if $p > -0.82$.\\

\item[(viii)] \textit{Symmetry statistic}: The second of the SPA criteria considered in this work, the symmetry statistic, $s$, measures the symmetry of a series of isophote ellipses about a global centre.
Isophote ellipses are fit to all pixels with a surface brightness between $S_{j}$ and $S_{j+1}$, with the centre, major and minor axes, and pitch left as free parameters.
The symmetry statistic is then defined as
\begin{equation}
 s=-\log_{10}\left(\frac{1}{N_{\mathrm{el}}}\sum_{j=1}^{N_{\mathrm{el}}}\frac{\delta_{j,c}}{\langle b_{\mathrm{el}}\rangle_{j}}\right)\:,
\end{equation}
where $N_{\mathrm{el}}$ is the number of ellipses fit\footnote{Note it is not always possible to fit all $5$ ellipses for a given cluster.}, $\delta_{j,c}$ is the $j$th ellipse and the globally determined centre and $b_{\mathrm{el}}$ is the average of the major and minor axes for the $j$th ellipse.
Following \citet{Mantz2015}, a cluster is classified as relaxed if $s > 0.87$.\\

\item[(ix)] \textit{Alignment statistic}: The final SPA criterion considered in this work, the alignment statistic, $a$, is sensitive to the presence of substructure emission at larger radii, which shifts the centre of emission for different isophote levels.
The alignment statistic is defined as
\begin{equation}
 a=-\log_{10}\left(\frac{1}{N_{\mathrm{el}}-1}\sum_{j=1}^{N_{\mathrm{el}}-1}\frac{\delta_{j,\,j+1}}{\langle b\rangle_{j,\,j+1}}\right)\:,
\end{equation}
where $\delta_{j,\,j+1}$ is the distance between the centres of the $j$th and $(j\!+\!1)$st isophote ellipses and $\langle b\rangle_{j,\,j+1}$ is the average of the four (two major, two minor) ellipse axes lengths.
Clusters are classified as relaxed by the alignment statistic if $a > 1.00$ \citep{Mantz2015}.\\
\end{itemize}

Like the theoretical morphological parameters, we refer to the centroid shift, power ratio and photon asymmetry statistic as ``negative'' parameters, i.e. a smaller value implies a more relaxed cluster.
However, the SPA criteria are ``positive'' parameters, where a larger value implies a more relaxed cluster.
Therefore, in all figures throughout this work, we invert the SPA parameter axes to ensure that clusters classified as relaxed appear on either the left or bottom of the axes for all criteria.
In Table \ref{tab:rlx_params}, we summarize the $9$ criteria examined in this work, the threshold value chosen and the literature reference for that threshold.

\begin{table}
 \caption{Table summarizing the morphological criteria examined in this work. The columns denote the parameter name, its symbol, the selected threshold for classifying clusters as relaxed and the literature reference for the chosen threshold.}
 \centering
 \begin{tabular}{Z{2.36cm} Y{0.85cm} Y{1.1cm} Z{2.51cm}}
  \hline
  Parameter & Variable & Threshold & Literature \\
  \hline
  Centre of mass offset & $X_{\mathrm{off}}$ & $< 0.07$ & \citet{Neto2007} \\
  Substructure fraction & $f_{\mathrm{sub}}$ & $< 0.1$ & \citet{Neto2007} \\
  Energy ratio & $E_{\mathrm{rat}}$ & $< 0.1$ & \citet{Barnes2017a} \\
  Centroid shift & \cshift & $< 0.006$ & \citet{Maughan2012} \\
  Power ratio & $P_{3}/P_{0}$ & $< 10^{-8}$ & \citet{Rasia2013} \\
  Photon asymmetry & $A_{\mathrm{phot}}$ & $< 0.15$ & \citet{Nurgaliev2017} \\
  Peakiness statistic & $p$ & $> -0.82$ & \citet{Mantz2015} \\
  Symmetry statistic & $s$ & $> 0.87$ & \citet{Mantz2015} \\
  Alignment statistic & $a$ & $> 1.00$ & \citet{Mantz2015} \\
  \hline
 \end{tabular}
 \label{tab:rlx_params}
\end{table}

\begin{figure*}
  \centering
  \includegraphics[width=\textwidth]{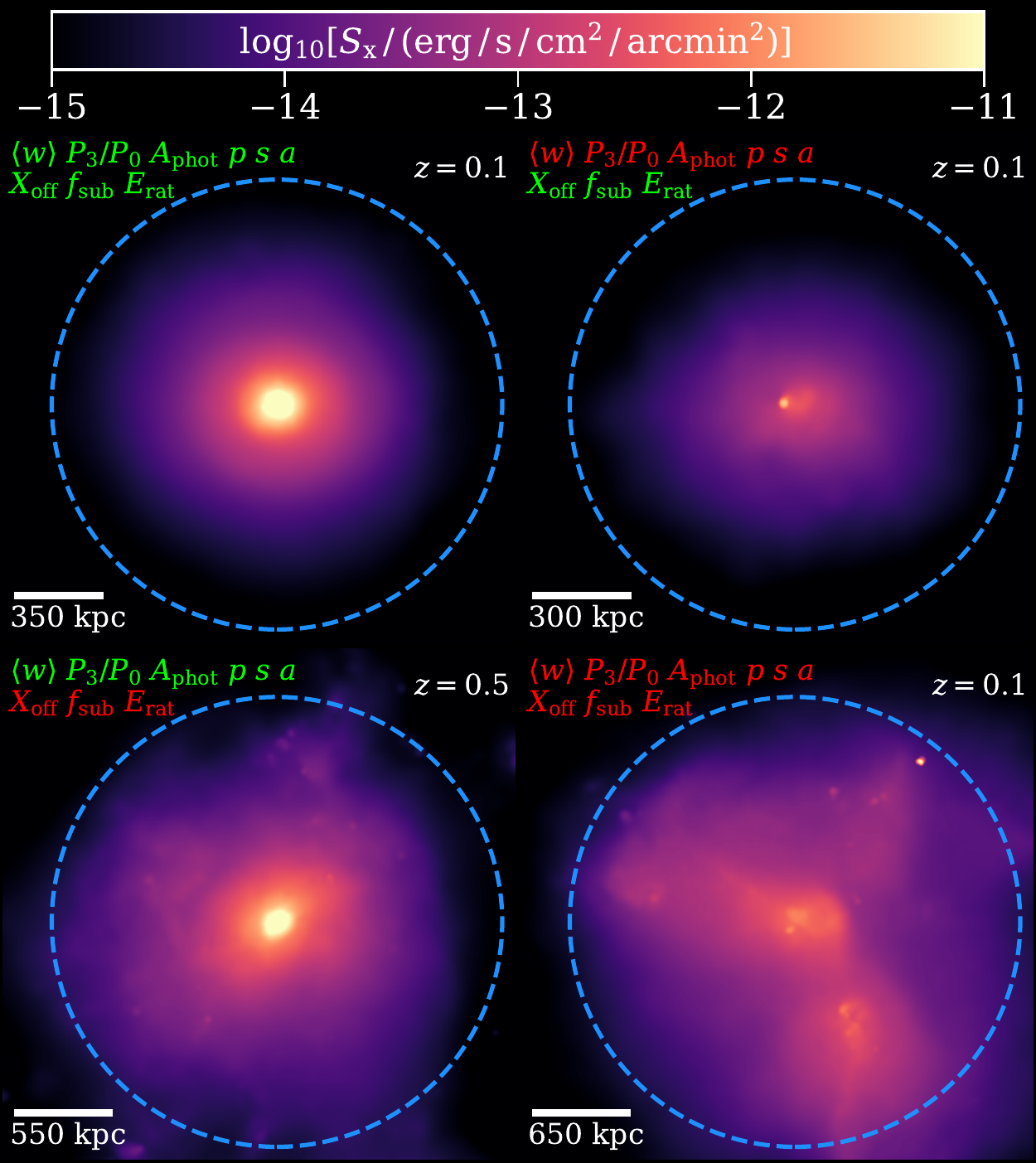}
  \caption{Four exemplar smoothed surface brightness maps computed from synthetic X-ray images of simulated clusters that pass all (top left), the theoretical (top right), the observational (bottom left) and none (bottom right) of the relaxation criteria examined in this work. The dashed blue line denotes $r_{\mathrm{500,crit}}$ of the cluster. Observational metrics focus on strong central emission and a smooth azimuthal distribution of emission, but theoretical parameters appear more sensitive to structure within the ICM.}
  \label{fig:exemplar_maps}
\end{figure*}

Figure \ref{fig:exemplar_maps} demonstrates the morphological features the $9$ criteria focus on. 
Each of the four panels presents a smoothed surface brightness image of a cluster taken from either the BAHAMAS or MACSIS samples.
The top left panel highlights a cluster that is defined as relaxed by all criteria explored in this work.
The cluster has a regular, circular X-ray emission with a strong central emission and would very likely be classified as relaxed by visual classification.
The top right panel shows a cluster classified as relaxed by the theoretical criteria, but it is not defined as relaxed by the observational metrics.
The cluster lacks the strong central emission of the previous image and the emission is significantly less spherical, but the X-ray emission has no obvious signs of substructures.
The bottom left panel displays a cluster that is classed as relaxed by the observational criteria, but not the theoretical metrics.
The contrast of these images demonstrates that the theoretical criteria appear to be more focused on the presence of structure within the ICM, even if the X-ray emission associated with them is relatively smooth.
For the observational metrics, strong central emission and relatively even azimuthal distribution of emission appears to be more important.
The bottom right panel highlights a synthetic image that is classified as unrelaxed by all criteria, and it is disturbed with asymmetric emission, lacks strong central emission and shows clear evidence of significant structure within the ICM.

\subsection{Statistical techniques}
\label{sec:stat_techs}
We now outline the statistical methods used in this paper.
The CDF fitting method and correlation measures are summarized in Sections \ref{sec:CDF_fit} and \ref{sec:corr_coefs}, respectively.

\subsubsection{Cumulative distribution fitting}
\label{sec:CDF_fit}

\begin{figure}
  \centering
  \includegraphics[width=\columnwidth, keepaspectratio=True]{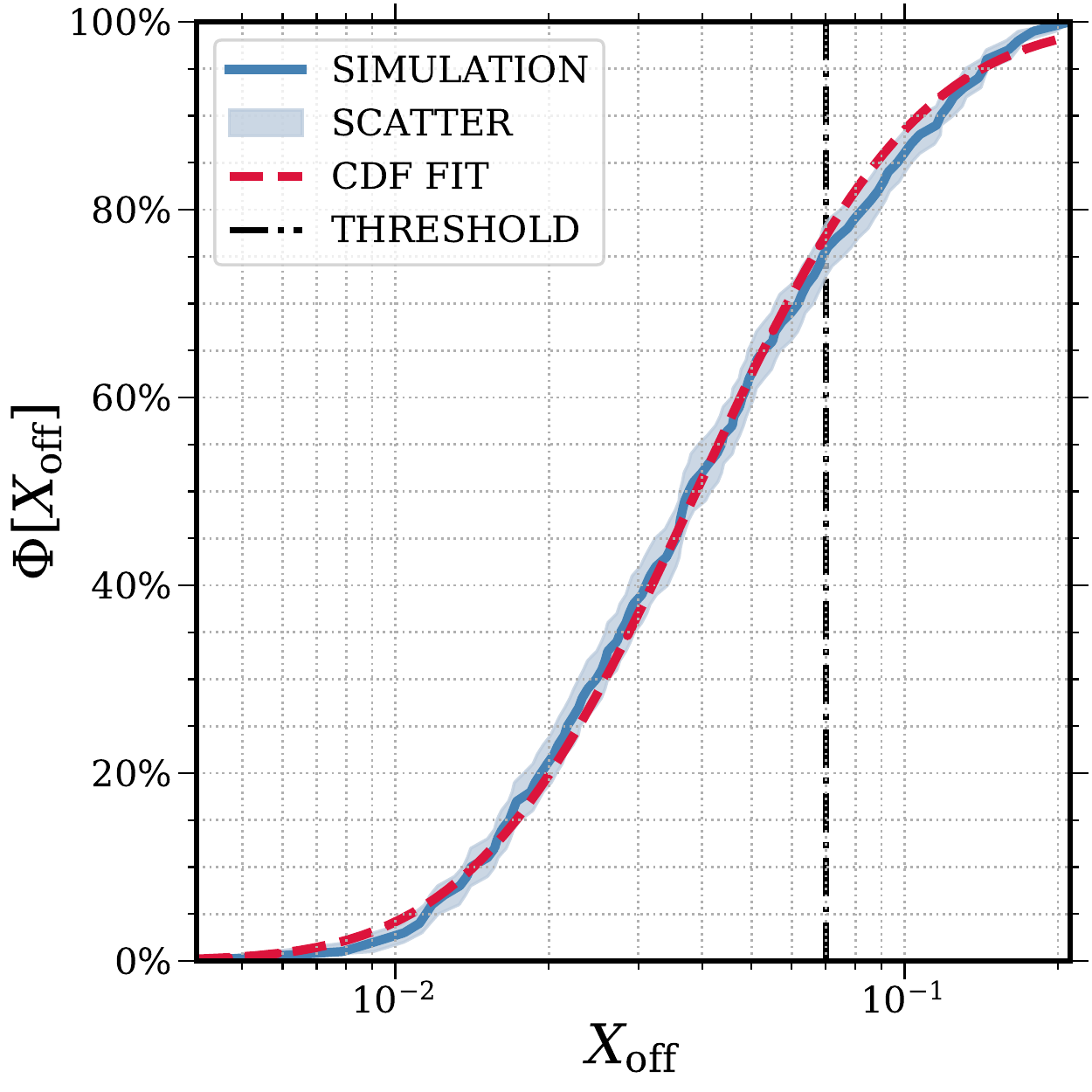}
  \caption{Log-normal distribution and CDF fit of the centre of mass displacement morphological metric, $X_{\mathrm{off}}$, for the TNG300-L1 sample at $z = 0.1$. The solid blue and dashed red lines show the sample distribution and best fit, respectively. The shaded region denotes the $1\sigma$ uncertainty computed via $10,000$ boostrap resamples. The dash-dot black line illustrates the literature threshold \citep{Neto2007}.}
 \label{fig:CDF_fit_diagram}
\end{figure}

As demonstrated in Figure \ref{fig:CDF_fit_diagram}, we find that, with sufficient statistics, the cumulative distribution functions (CDFs) of all morphological metrics, both observational and theoretical, are well described by log-normal distributions.
We note that the SPA criteria by definition include the logarithm that converts them to Gaussian distributions.
For some samples with small number statistics, such as the TNG300 samples, the distribution can be significantly noisy; and for some parameters, such as the power ratio $(P_{3}/P_{0})$, the distribution can have a significant tail.
To reduce the impact of noise and extreme outliers in some of the analyses presented below, we make use the of the best-fit distribution to the cumulative distribution function of every parameter at every redshift for all samples.
In the limit of a statistically large sample with few extreme outliers, the results covered from the best-fit are identical to those yielded by other techniques, such as maximum likelihood estimation.
All fitting parameters are given in the tables of Appendix \ref{app:dist}. 

We compute the best fit Gaussian CDF by performing a least-squares fit that minimizes
\begin{equation}\label{eq:pcnt_cdf}
 \chi^2 = \sum_{i=1}^{99}\left[\frac{\mu+\sigma\sqrt{2}\,\mathrm{erf}^{-1}(2\Phi_i-1)-\mathrm{pcnt}_i}{\mathrm{std}_i}\right]^2\:,
\end{equation}
where $\mu$ and $\sigma$ are the usual mean and standard deviation of a Gaussian function, $\mathrm{erf}^{-1}$ is the inverse error function, $\Phi_i \equiv i / 100$ in our case is the cumulative distribution, $\mathrm{pcnt}_i$ is the value of the sample CDF at the $i$th percentile and $\mathrm{std}_i$ is the uncertainty in the CDF at the $i$th percentile computed via $10,000$ bootstrap resamples.
The fit is performed between the $1$st and $99$th percentiles to reduce the impact of extreme outliers and to avoid the inverse error function tending to infinity.

\subsubsection{Correlation coefficients}
\label{sec:corr_coefs}
We measure the correlation between the morphological parameters using two different coefficients
\begin{itemize}
 \item[(i)] \textit{Pearson correlation coefficient}: Measures the linear correlation between to variables via
 \begin{equation}
  r_{\mathrm{p}} = \frac{cov(X,Y)}{\sigma_{\mathrm{X}}\sigma_{\mathrm{Y}}}\:.
 \end{equation}
 \item[(ii)] \textit{Spearman's rank coefficient}: A non-parametric measure of rank correlation between two variable
 \begin{equation}
  r_{\mathrm{s}}=r_{\mathrm{p,g_{\mathrm{X}},g_{\mathrm{Y}}}}\:,
 \end{equation}
 where $r_{\mathrm{p}}$ is the Pearson correlation coefficient applied to the ranked variables $g_{\mathrm{X}}$ and $g_{\mathrm{Y}}$.
\end{itemize}
If the distributions are truly log-normal then these two correlation measures should return very similar values.
All correlation coefficients are always computed for variables that are normally distributed, i.e. we log all criteria that do not include it by construction include it.
For the ``positive'' parameters, i.e. the SPA criteria, we invert the distribution to ensure positive correlations with the ``negative'' criteria.

\begin{figure*}
 \centering
 \includegraphics[width=\textwidth]{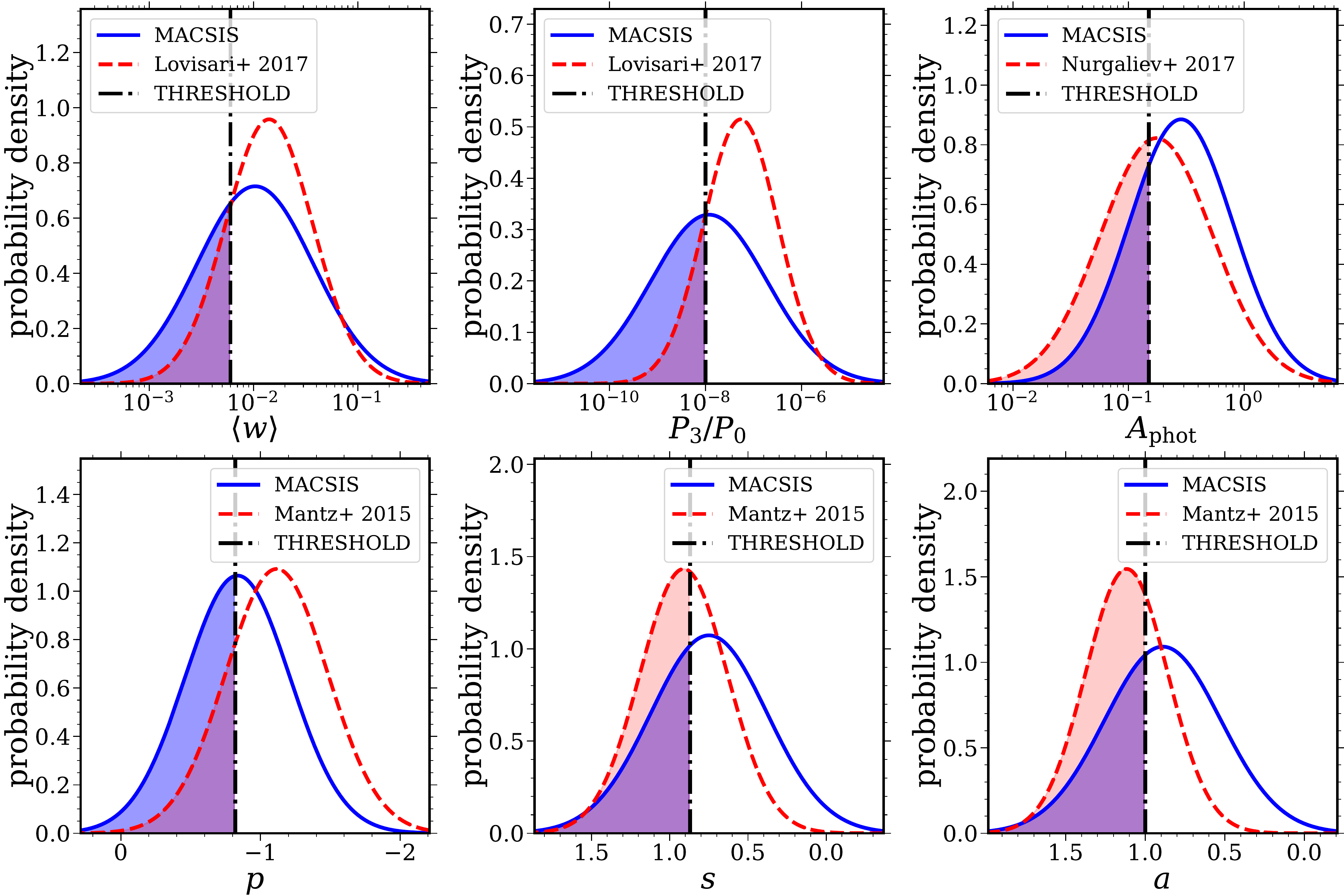}
 \caption{Comparison between simulated (solid blue) and observed (dashed red) morphological criteria distributions. The observed distributions are extracted from \citet{Mantz2015}, \citet{Lovisari2017}, and \citet{Nurgaliev2017}. The simulated samples are cut to ensure the median mass/temperature and redshift are well matched to the observed sample. The blue (red) shaded region denotes the fraction of the distribution that would be classified as relaxed. The black dash-dot line denotes the threshold value taken from the literature. We note that the SPA criteria axes are inverted such that relaxed clusters always appear on the left of the panel.}
 \label{fig:dist_CF}
\end{figure*}

\section{Criteria distributions}
\label{sec:dist}
We now examine the distributions of the morphological metrics outlined in Section \ref{sec:rlx_crit}, exploring their evolution with mass and redshift, and comparing to observed distributions.
In order to reduce the impact of noise due to small sample sizes, we fit CDF distributions, as outlined in Section \ref{sec:CDF_fit}, to both the simulated and observed samples.

\subsection{Comparison with observed distributions}
\label{sec:dist_CF}

We begin by comparing the observed distributions to matched simulated distributions.
We extract observational data from \citet{Lovisari2017} (\cshift, $P_{3}\,/\,P_{0}$), \citet{Nurgaliev2017} ($A_{\mathrm{phot}}$) and \citet{Mantz2015} ($s$, $p$ and $a$).

\citet{Lovisari2017} analyzed the \textit{Planck} early Sunyaev-Zeldovich (ESZ) cluster sample \citep{Planck2011}, which consists of $120$ massive clusters with a median mass $M_{500}=6.1\times10^{14}\,\mathrm{M}_{\astrosun}$.
Due to the relatively high median mass of the sample, with SZ surveys yielding mass-selected-\textit{like} samples \citep{Lin2015,Mantz2019}, we compare to a simulated sample extracted from the MACSIS sample at $z=0.1$ and $0.3$.
We make use of mass estimated from the synthetic X-ray images, assuming hydrostatic equilibrium \citep[see][for further details]{Barnes2020}, to better match the observed sample, whose masses are derived from the $M-Y_{\mathrm{X}}$ relation of \citet{Arnaud2007} which is calibrated with hydrostatic mass estimates.
We apply a mass cut of $M_{\mathrm{500,X\mbox{-}ray}}=2.7\times10^{14}\,\mathrm{M}_{\astrosun}$ to the MACSIS sample, which yields a sample of $3856$ clusters with a median $M_{500}=6.1\times10^{14}\,\mathrm{M}_{\astrosun}$.

\citet{Nurgaliev2017} studied $90$ clusters from the 2500 deg$^2$ South Pole Telescope (SPT) survey \citep{Bleem2015} and $36$ high-$z$ clusters from the ROSAT PSPC 400 deg$^2$ cluster survey \citep{Burenin2007}.
The median mass of this sample is $M_{\mathrm{500}}=4.1\times10^{14}\,\mathrm{M}_{\astrosun}$.
Due to the high-redshift and massive nature of the sample, the MACSIS sample is the appropriate comparison.
We select clusters from the $z=0.3$, $0.5$ and $1.0$ outputs and make a mass cut at $M_{\mathrm{500,X\mbox{-}ray}}=1.8\times10^{14}\,\mathrm{M}_{\astrosun}$, to yield a simulated sample of $5042$ clusters with a median mass of $M_{\mathrm{500}}=4.1\times10^{14}\,\mathrm{M}_{\astrosun}$.

Finally, \citet{Mantz2015} studied $362$ clusters extracted from the \textit{Chandra} archive with a a median core-excised temperature of $kT_{\mathrm{X,ce}}=6.8\,\mathrm{keV}$.
Given the high median temperature of the sample, the MACSIS clusters are the correct comparison and we select $2794$ simulated haloes from the $z=0.1$, $0.3$ and $0.5$ snapshots using a temperature cut of $kT_{\mathrm{X,ce,500,X\mbox{-}ray}}=5.5\,\mathrm{keV}$.
This produces a simulated sample with a median core-excised temperature $kT_{\mathrm{X,ce,500,X\mbox{-}ray}}=6.8\,\mathrm{keV}$.
For a detailed explanation of how we compute the core-excised spectroscopic temperature from the synthetic X-ray image, we refer the reader to Barnes et al. (in prep.).

Figure \ref{fig:dist_CF} compares the observed and simulated X-ray morphological criteria distributions.
The simulated centroid shift (\cshift) distribution yields a slightly more relaxed distribution than the observed sample from \citet{Lovisari2017}, with mean values of $\log_{10}\cshift=-1.985\pm0.007$ and $-1.851\pm0.004$, respectively.
Additionally, the simulated distribution is wider with $\sigma=0.558\pm0.007$ compared to $0.416\pm0.004$ to the observed.
Consequently, the fraction of clusters classified as relaxed in the simulated sample ($35$ per cent) is significantly larger than that of the observed sample ($21$ per cent).
Though not perfect, we find a reasonable overlap between the simulated and observed distributions.
The discrepancies between them are likely the combination of unaccounted for selection effects and the limitations of the subgrid model.

The third-order power ratio shows a similar result, with the simulated distribution having a smaller mean value ($\log_{10}(P_{3}/P_{0})=-7.927\pm0.012$ versus $-7.264\pm0.006$) and larger width ($1.213\pm0.015$ versus $0.775\pm0.009$) relative to the observed sample.
This leads to a significantly larger fraction of simulated clusters being classified as relaxed ($49$ per cent) relative to the observed sample ($17$ per cent).
The power ratio shows the largest difference between the simulated and observed distributions.
However, there is still reasonable overlap between the distributions, suggesting that the simulated distribution is still a plausible representation of observed clusters. 

The photon asymmetry metric yields two samples whose distributions are relatively similar, with the simulated and observed samples having very similar widths, $0.451\pm0.003$ and $0.485\pm0.008$ respectively, and a minor difference in mean value, $\log_{10}A_{\mathrm{phot}}=-0.546\pm0.003$ and $-0.759\pm0.007$ respectively.
Adopting the relaxation threshold from \citet{Nurgaliev2017}, $A_{\mathrm{phot}}<0.15$, we find that $29$ per cent and $48$ per cent of the simulated and observed clusters, respectively, are defined as relaxed.
Although the simulated and observed distributions show significant overlap, the $\sim60$ per cent increase in the fraction of clusters classified as relaxed highlight how sensitive the relaxed fraction can be to the chosen threshold.

Comparison to the peakiness ($p$), symmetry ($s$) and alignment ($a$) morphological metrics is slightly more complicated.
The other observational samples are drawn from SZ samples, which are less sensitive to selection effects relative to the \citet{Mantz2015} sample drawn primarily from the \textit{Chandra} archive data of previous massive cluster surveys.
Therefore, we try to mimic this selection by using the MACSIS cluster sample and matching the measured X-ray temperature.
However, we note that we make no other attempt to match the likely highly complex selection function present in the observed sample.
We find simulated peakiness, symmetry and alignment parameter mean values of $p=-0.836\pm0.012$, $s=0.749\pm0.003$ and $a=0.890\pm0.003$.
In comparison, the observed distributions yield mean values of $p=-1.116\pm0.005$, $s=0.911\pm0.002$ and $a=1.117\pm0.002$, which yields a greater fraction of relaxed clusters for symmetry and alignment, but a smaller relaxed fraction for the peakiness metric.
Additionally, the simulated distributions are again wider than the observed distributions for the alignment and symmetry parameters, with the peakiness distributions of similar width.
However, there is still reasonable overlap between the simulated and observed samples and we believe the simulations yield reasonable distributions for the SPA morphological metrics.

We conclude that although the simulated and observed samples are not perfectly matched, there is reasonable agreement between them.
Part of the reason that the samples are not perfectly matched may be selection effects, though we match the median mass of the observed sample we do not mimic anything else about the sample selection or the impact Malmquist bias and other effects.
A further investigation of these effects will require synthetic surveys, rather than images, and we will revisit these effects in future work.
Having found that the simulated distributions are not unreasonable, we now explore how the morphological metric distributions evolve with redshift for the different simulated samples.

\subsection{Redshift evolution}
\label{sec:dist_sim}

\begin{figure*}
 \includegraphics[width=\textwidth]{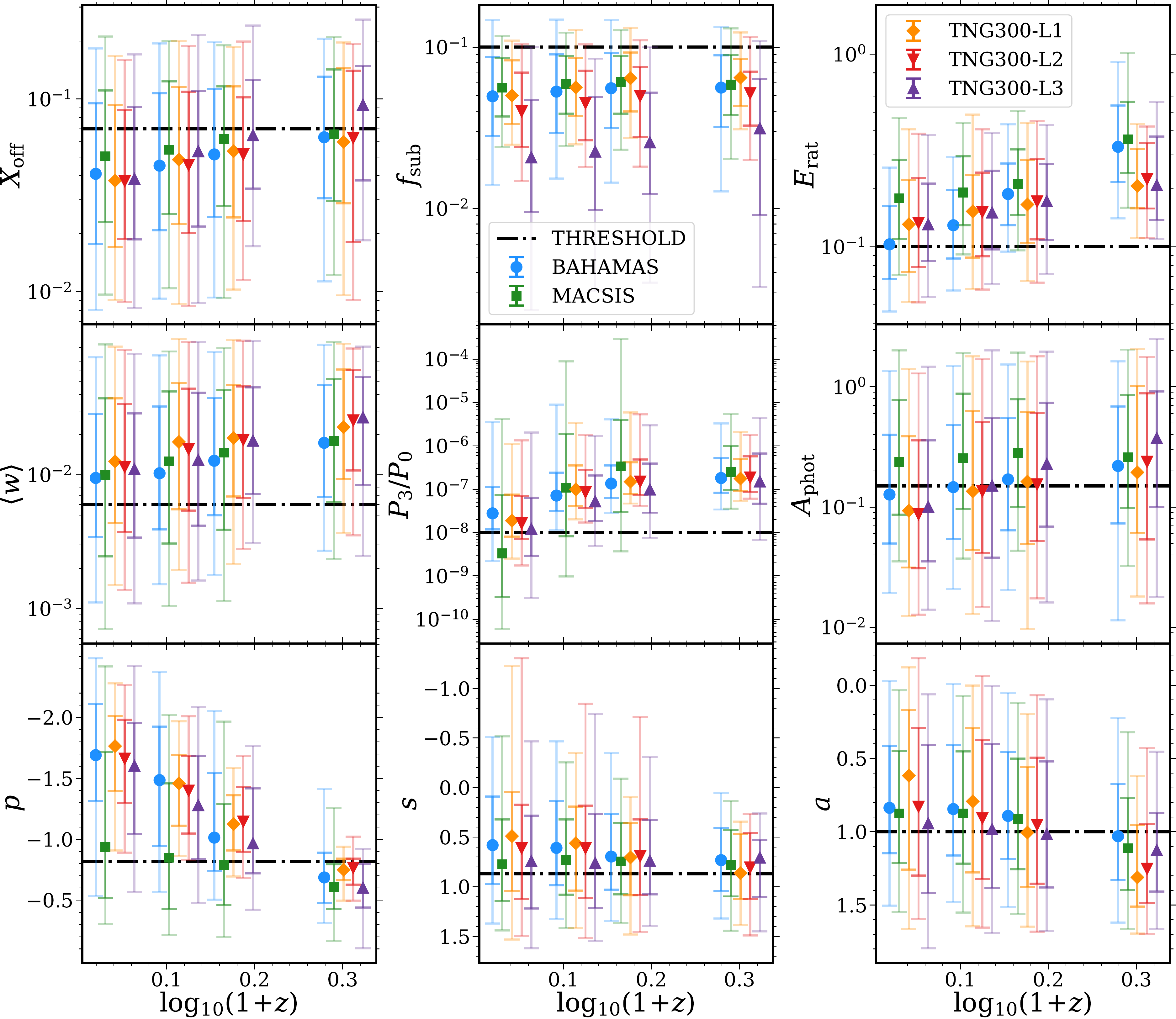}
 \caption{Redshift evolution of the morphological criteria examined in this work. Cluster samples from TNG300 level 1 (orange diamond), level 2 (red down triangle), level 3 (purple upward triangle), BAHAMAS (blue circle) and MACSIS (green square) are shown. The error bars show $68$ and $95$ percent of the sample. The black dash-dot shows the literature threshold value. For clarity, we have introduced small redshift offsets between the points.}
 \label{fig:dist_pcnts}
\end{figure*}

Figure \ref{fig:dist_pcnts} shows the redshift evolution of the $9$ morphological criteria explored in this work for the $5$ simulated cluster samples and we now discuss each metric in turn.
Where appropriate, we simultaneous fit the $5$ simulated samples with the functional form
\begin{equation}
 \log_{10}(y)=K+\beta\log_{10}(1+z)\:,
\end{equation}
to determine the average redshift evolution of a given criterion.
Thus, we are then able to remove it by incorporating $(1+z)^{-\beta}$ in the morphological metric.
We remind the reader that the peakiness criteria already includes a $(1+z)$ term.\\ 

\noindent\textit{Centre of mass offset}: The top left panel of Figure \ref{fig:dist_pcnts} shows the redshift evolution of the centre of mass offset.
There is a clear redshift evolution trend in the criterion for all simulated samples, with the average median value increasing from $4.10\times10^{-2}$ to $6.90\times10^{-2}$ from $z=0.1$ to $z=1.0$.
This evolution is most likely driven by the fact the merger rate of haloes increases with redshift \citep[e.g.][]{Carlberg1997,Nelson2014} and that clusters of a fixed mass at $z=1.0$ have had significantly less time to relaxed relative to their low-redshift counterparts \citep[e.g.][]{Kunz2011,Zhuravleva2014}.
Fitting the evolution, we find a best-fitting value $\beta=0.86\pm 0.13$ must be applied if a single value threshold is to be used for all redshifts.
The MACSIS sample at low-redshift $(z=0.1)$ has a larger median $X_{\mathrm{off}}=5.05\times10^{-2}$ than the BAHAMAS sample $X_{\mathrm{off}}=4.09\times10^{-2}$, and this continues up to $z=0.5$.
This is likely the result of more massive clusters having formed more recently and had less time to thermalize, with the MACSIS and BAHAMAS samples having a median mass $M_{\mathrm{500,sim}}=7.83\times10^{14}\,\mathrm{M}_{\astrosun}$ and $M_{\mathrm{500,sim}}=1.10\times10^{14}\,\mathrm{M}_{\astrosun}$, respectively.
At $z=1.0$, the difference between the two samples disappears as our mass selection threshold yields samples with similar median mass, $M_{\mathrm{500,sim}}=2.00\times10^{14}\,\mathrm{M}_{\astrosun}$ and $M_{\mathrm{500,sim}}=0.89\times10^{14}\,\mathrm{M}_{\astrosun}$ for MACSIS and BAHAMAS, respectively.
At low-redshift $(z\leq0.5)$, with sufficiently large sample sizes, the BAHAMAS and TNG300 level 1 samples yield consistent median values, suggesting the subgrid model has a minimal impact on the metric.
However, there does appear to be some numerical resolution dependence, with the TNG300 level 3 sample yielding a median value that is $56$ per cent larger than the level 1 sample at $z=1.0$, respectively.
The difference due to numerical resolution decreases towards low redshift, but we require larger samples to draw definite conclusions regarding this dependence.\\

\noindent\textit{Substructure mass fraction}: The top middle panel of Figure \ref{fig:dist_pcnts} shows that the substructure fraction is relatively independent of redshift, with the best fit value $\beta=0.38\pm 0.36$ consistent with negligible evolution.
However, there is significant evidence that it depends on the numerical resolution of the simulated sample.
At fixed cluster mass, a higher mass resolution enables smaller substructures within the cluster volume to be resolved and a greater fraction of the cluster mass becomes bound to substructures.
This becomes clear for the TNG300 samples, the level 1 sample median substructure mass function is $25$ and $142$ per cent larger than the level 2 and 3 samples, respectively, at $z=0.1$.
This difference remains relatively constant with redshift.
Due to the self-similar nature of large-scale structure formation, at a fixed numerical resolution more substructures are resolved in more massive clusters because the substructures are more massive.
The BAHAMAS and MACSIS samples highlight this effect, with the MACSIS sample substructure fraction $13$ per cent larger than the BAHAMAS sample at $z=0.1$.
With increasing redshift, the offset between the two samples is removed as the difference in the median mass of the samples reduces due to our cluster selection by a mass threshold.
The substructure mass fraction is unique among the morphological criteria considered in this study, as it is the only one to consistently classify the majority of clusters as relaxed.
This highlights theoretical criteria typically setting thresholds to remove the most disturbed objects rather than selected a subset of the most relaxed objects.\\

\noindent\textit{Energy ratio}: The energy ratio shows the strongest redshift evolution of the theoretical criteria, as shown in the top right panel of Figure \ref{fig:dist_pcnts}.
As above, this is because the merger rate of haloes increases with redshift and they have had less time to convert the kinetic energy of infalling structures and material into thermal energy.
More massive, hence more recently formed, clusters yield a larger value of the energy ratio, as shown by MACSIS sample producing a median $E_{\mathrm{rat}}$ value $73$ per cent larger than the BAHAMAS sample at $z=0.1$.
Though increasingly noisy at high redshift, the three numerical resolution levels of TNG yield consistent results, suggesting resolution has little impact on the energy ratio criteria.
The redshift evolution is best fit a value of $\beta=1.15\pm 0.18$.
Interestingly, the energy ratio highlights the differences between the subgrid models.
The TNG level 1 sample produces a larger $E_{\mathrm{rat}}$ value at low redshift than the BAHAMAS sample, but smaller values at high redshift.
This may be due to differences in the implementation of feedback in the two models and how this feedback generates bulk and turbulent motions in the ICM.
However, we leave a detailed investigation to future work.\\

\noindent\textit{Centroid shift}: As shown in the centre-left panel of Figure \ref{fig:dist_pcnts}, the centroid shift shows significant redshift evolution, driven by increased mergers and a shorter period over which to thermalize.
The redshift evolution fit yields $\beta=1.17\pm 0.17$.
The centroid shift appears to be relatively insensitive to numerical resolution, with all three TNG resolutions yielding similar results.
However, the BAHAMAS sample yields a median centroid shift that is $25$ per cent lower than the TNG300 level 1 sample at $z=0.1$, and this offset remains similar with increasing redshift.
This may be driven by how the cluster ICM reacts to both internal and external events with different hydrodynamics methods, but we are not able to draw definite conclusions.
With MACSIS and BAHAMAS producing similar values at all redshifts the centroid shift metric appears to be relatively insensitive to cluster mass.\\

\noindent\textit{Power ratio}: The power ratio is the only criterion that exhibits increasing evolution with decreasing redshift, as seen in the centre panel of Figure \ref{fig:dist_pcnts}.
Although the exact cause of this evolution is somewhat unclear, all five simulations demonstrate the same behaviour and it may be driven by substructure being more massive and, therefore, more luminous at lower redshifts, driving asymmetries in the photon distribution.
This would explain why the more massive on average MACSIS sample shows a greater degree of evolution relative to the other samples.
Though likely not a simple power-law, fitting for the redshift evolution yields a value of $\beta=4.07\pm 0.80$.
All TNG300 samples produce consistent median values at all redshifts, highlighting that the power ratio appears to be independent of numerical resolution.
Additionally, TNG300 level 1 and BAHAMAS yield similar median values, suggesting it is also insensitive to subgrid physics.\\

\noindent\textit{Photon asymmetry}: Shown in the centre right panel of \ref{fig:dist_pcnts}, The photon asymmetry statistic shows redshift evolution for all samples, with a best fit value of $\beta=1.26\pm 0.29$.
The is consistent with clusters being dynamical younger at higher redshifts due to increases mergers and shorter periods to thermalize.
We find that the criterion is mass-dependent, with the MACSIS sample yielding a median value that is $86$ per cent larger at $z=0.1$ than the BAHAMAS sample and the median MACSIS value evolving less with increasing redshift.
Though noisy at high redshift due to small sample sizes, the TNG300 samples and the BAHAMAS sample yield consistent median values, suggesting that photon asymmetry does not depend strongly on either numerical resolution or the subgrid physics model.\\

\noindent\textit{Surface brightness peakiness}: The surface brightness peakiness is the only morphological metric that evolves with increasing redshift to have a greater fraction of the sample defined as relaxed, as shown in the bottom right panel of Figure \ref{fig:dist_pcnts}.
The peakiness parameter already includes a $(1+z)$ term, and our result suggests this evolution, in the limit of perfect signal-to-noise, is too strong.
Finding the best-fit value of $\beta=3.42\pm 0.49$ for the redshift evolution, we tentatively suggest the actual redshift term should be $(1+z)^{-2}$.
However, this result is tempered by its dependence on one of the least certain aspects of galaxy formation models: AGN feedback.
The peakiness parameter is a measure of how centrally concentrated the X-ray emission is, which, due to the $n_{\mathrm{e}}^{2}$ dependence of the emission mechanism, is highly sensitive to conditions in the core of the cluster.
In numerical simulations, the primary mechanism for regulating the cores of galaxy clusters is AGN feedback.
Models of AGN feedback are often highly dependent on numerical resolution, as highlighted by the requirement (or lack thereof) of an accretion boost factor \citep[e.g.][]{Springel2005,BoothSchaye2009,Schaye2015}.
Additionally, the impact of AGN feedback changes with cluster mass, with the deeper potentials of more massive clusters reducing the impact of feedback.
We find the more massive MACSIS sample exhibits significantly less redshift evolution than the other samples and has a median value that is $44$ per cent smaller than the BAHAMAS median value at $z=0.1$.
Additionally, the BAHAMAS sample evolves more strongly with increasing redshift relative the TNG300 level 1 sample, highlighting that the parameter is sensitive to how the subgrid physics is implemented.
Finally, the parameter is also marginally dependent on the numerical resolution with the TNG300 level 1 sample yielding a median value that is $6$ and $10$ per cent larger than the level 2 and 3 samples, respectively, at $z=0.1$.
Therefore, though we find significant redshift evolution for the peakiness parameter any conclusion we can draw is limited by the current state of AGN feedback implementations in galaxy formation models.
It is highly likely that the explanation for the differences we find between the simulated samples is their implementation of AGN feedback and how it varies with numerical resolution and cluster mass.\\

\noindent\textit{Symmetry statistic}: The symmetry statistic, as shown in the lower centre panel of Figure \ref{fig:dist_pcnts}, already contains a $(1+z)$ term by construction.
However, we still find a mild redshift evolution, with a best-fit value of $\beta=0.59\pm 0.18$.
Additionally, there appears to be a mild dependence on mass and subgrid implementation, with the MACSIS (TNG300 level 1) sample returning a median symmetry value that is $33$ per cent larger ($16$ per cent smaller) than the BAHAMAS value at $z=0.1$.
These difference reduce with increasing redshift.
The width of the distribution reduces at $z=1.0$ relative to the other snapshots, however, this may simply be the result of small number statistics for this snapshot.\\

\noindent\textit{Alignment statistic}: By construction the alignment statistic contains a $(1+z)$ term, but, as shown in Figure \ref{fig:dist_pcnts}, we find a redshift evolution with a best-fit value $\beta=1.38\pm 0.21$.
The MACSIS and BAHAMAS samples produce consistent median values at all redshifts, suggesting that the statistic value does not depend on halo mass.
Though noisy, there does appear to be some dependence on subgrid physics and numerical resolution for the alignment statistic, with the TNG300 level 1 sample yielding a mildly different redshift evolution and median value relative to the other resolution levels and BAHAMAS.
However, the statistical uncertainty is relatively large and would require a significantly larger sample to investigate fully.\\

Overall, we find that all of the morphological metrics show some degree of redshift evolution, even when they include a $(1+z)$ term in their construction.
Therefore, the fraction of clusters that are classified as relaxed will evolve with redshift when using a fixed threshold value.
Given that all morphological metrics aim to describe some aspect of the dynamical nature of the ICM, one would expect the different criteria to be correlated at some level.
We now examine the extent of the correlation between them.

\section{Correlation}
\label{sec:corr}
We now examine the correlations between the morphological metrics.
First, we compare the simulated correlations to observed values published in the literature.
Then we explore the correlations between all criteria considered in this work, specifically comparing the different simulated samples to understand any evolution with mass, redshift or subgrid physics.

\subsection{Observational comparison}
\label{sec:corr_CF}

\renewcommand\arraystretch{1.5}
\begin{table}
 \caption{Comparison of simulated and observed morphological criteria correlations. The observational data is extracted from \citet{Mantz2015}, \citet{Lovisari2017}, and \citet{Nurgaliev2017}. Following Section \ref{sec:dist_CF}, the simulated (MACSIS) samples are selected to ensure the median mass/temperature and redshift are well matched to the observed sample.}
 \centering
 \begingroup
 \setlength{\tabcolsep}{3.8pt}
 \begin{tabular}{c|cc|cc}
  \hline
  Parameter pair & \multicolumn{2}{c|}{Observed sample} & \multicolumn{2}{c}{Simulated sample} \\
  \\[-2.5em]
  & $r_{\mathrm{p}}$ & $r_{\mathrm{s}}$ & $r_{\mathrm{p}}$ & $r_{\mathrm{s}}$ \\
  \hline
  $\langle w \rangle$ and $P_{3}/P_{0}$ & $0.530^{+0.065}_{-0.072}$ & $0.570^{+0.062}_{-0.067}$ & $0.434^{+0.013}_{-0.013}$ & $0.414^{+0.014}_{-0.013}$ \\
  $\langle w \rangle$ and $A_\mathrm{phot}$ & $0.751^{+0.036}_{-0.045}$ & $0.739^{+0.041}_{-0.041}$ & $0.852^{+0.004}_{-0.005}$ & $0.832^{+0.004}_{-0.004}$ \\
  $\langle w \rangle$ and $p$ & $0.561^{+0.049}_{-0.055}$ & $0.565^{+0.050}_{-0.052}$ & $0.646^{+0.011}_{-0.012}$ & $0.629^{+0.010}_{-0.010}$ \\
  $\langle w \rangle$ and $s$ & $0.765^{+0.027}_{-0.032}$ & $0.719^{+0.046}_{-0.034}$ & $0.774^{+0.008}_{-0.009}$ & $0.737^{+0.010}_{-0.010}$ \\
  $\langle w \rangle$ and $a$ & $0.670^{+0.039}_{-0.043}$ & $0.642^{+0.046}_{-0.041}$ & $0.711^{+0.010}_{-0.010}$ & $0.691^{+0.010}_{-0.010}$ \\
  $p$ and $s$ & $0.450^{+0.050}_{-0.055}$ & $0.415^{+0.053}_{-0.056}$ & $0.604^{+0.013}_{-0.013}$ & $0.664^{+0.012}_{-0.012}$ \\
  $p$ and $a$ & $0.435^{+0.053}_{-0.054}$ & $0.367^{+0.068}_{-0.063}$ & $0.557^{+0.014}_{-0.014}$ & $0.615^{+0.013}_{-0.013}$ \\
  $s$ and $a$ & $0.724^{+0.030}_{-0.034}$ & $0.710^{+0.029}_{-0.028}$ & $0.801^{+0.008}_{-0.008}$ & $0.800^{+0.007}_{-0.007}$ \\
  \hline
 \end{tabular}
 \endgroup
 \label{tab:corr_CF}
\end{table}
\renewcommand\arraystretch{1.0}

Table \ref{tab:corr_CF} compares the simulated morphological metric correlations to those extracted from the literature and presents the uncertainties on the measured coefficients.
Following Section \ref{sec:dist_CF}, we compare to data extracted from \citet{Lovisari2017} (\cshift, $P_{3}\,/\,P_{0}$), \citet{Nurgaliev2017} (\cshift, $A_{\mathrm{phot}}$) and \citet{Mantz2015} (\cshift, $s$, $p$ and $a$).
Note that we extract centroid shift for all three observational samples.
We again select the MACSIS sample as the appropriate comparison and apply the appropriate mass/temperature cut to ensure that the median of the simulated sample is well matched to the observational sample.
However, we again highlight that we make no further attempts to match the selection functions of the observed samples.

We find good general agreement between the correlations of the observed morphological metrics and those produced by the simulated sample, with those metrics that are observed to more tightly correlated also more strongly correlated in the simulated samples.
For example, measuring the Pearson correlation coefficient between centroid shift and the power ratio for the \textit{Planck} ESZ sample \citep{Lovisari2017} yields a value of $0.53$, which is smaller than the value of $0.75$ produced by the photon asymmetry statistic and the centroid shift for the SPT cluster sample \citep{Nurgaliev2017}.
The simulated samples produce values of $0.434$ and $0.852$ for the centroid shift-power ratio and centroid shift-photon asymmetry parameter correlations, respectively.

The centroid shift is also observed to be reasonably well correlated with the SPA criteria, yielding Pearson coefficients of $0.561$, $0.765$ and $0.670$ for the peakiness, symmetry and alignment parameters, respectively.
The simulated sample produces correlation coefficients in good agreement with the observations, with values of $0.646$, $0.774$ and $0.711$.
Finally, for the SPA criteria themselves the observational data yields correlations of $0.450$, $0.435$ and $0.724$ for the peakiness-symmetry, peakiness-alignment and symmetry-alignment criteria combinations, respectively.
The simulated sample produces marginally stronger correlation than the observational samples with $0.604$, $0.557$ and $0.801$ for the same metric combinations.

In general, we find that simulated samples yield similar correlation trends to the observational data.
However, for some statistics, the magnitude of the correlation is mildly stronger for the simulated samples relative to the observational data.
This is potentially linked to the lack of noise in our synthetic images, but the differences may also be linked to selection effects that we have not modelled in this analysis.
A detailed study of the impact of both these effects is beyond the scope of this paper as it will require synthetic surveys.
We conclude that the simulations yield reasonable correlation values between the different morphological metrics and we now examine their evolution with redshift, mass and subgrid physics.

\subsection{Simulation results}
\label{sec:corr_sim}

\begin{figure*}
 \includegraphics[width=\textwidth]{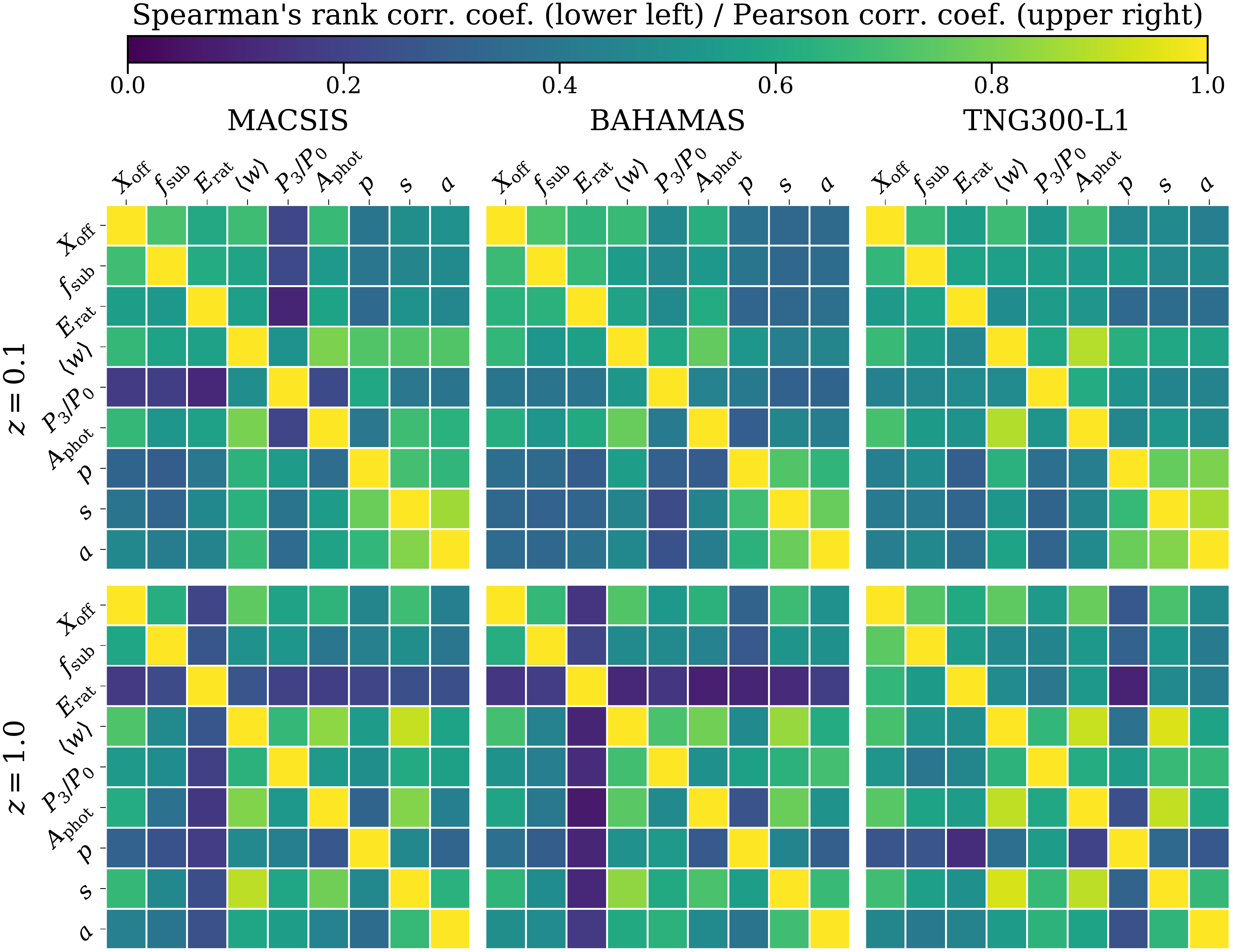}
 \caption{Correlation coefficient matrices for the $9$ morphological metrics at $z=0.1$ (top row) and $z=1.0$ (bottom row). Results for the MACSIS (left), BAHAMAS (middle) and TNG300 level 1 (right) samples are presented. Pearson correlation coefficients (Spearman's rank correlation coefficients) are shown in the lower left (upper right) triangle area. We note that we invert the values of the ``positive'' SPA criteria to ensure that if both criteria predict a more relaxed cluster the correlation is positive.}
 \label{fig:corr}
\end{figure*}

Figure \ref{fig:corr} shows the Pearson and Spearman's rank correlation coefficients for the MACSIS, BAHAMAS and TNG300 level 1 samples at $z=0.1$ and $z=1.0$.
We have examined all samples at all redshifts, but we highlight these samples at the two redshift extremes to demonstrate how the correlations between the morphological metrics evolve with mass, redshift and subgrid physics.
Appendix \ref{app:corr} shows the complete correlation matrices for all samples and redshifts.

At $z=0.1$, the theoretical criteria are significantly correlated with each other for all three samples, with coefficient values in the range $0.6-0.7$.
However, at $z=1.0$, we find that these correlations weaken substantially for the MACSIS and BAHAMAS samples.
For example, the correlation between the energy ratio, $E_{\mathrm{rat}}$ and both the centre of mass offset, $X_{\mathrm{off}}$, and substructure mass fraction, $f_{\mathrm{sub}}$, drops to $\lesssim 0.3$ and $\lesssim 0.2$ in the MACSIS and BAHAMAS samples, respectively.
The TNG300 level 1 sample, on the other hand, yields a stronger correlation between $X_{\mathrm{off}}$ and $f_{\mathrm{sub}}$ at $z=1.0$, with Pearson and Spearman's rank values of $0.789$ and $0.782$, respectively.
Additionally, the $E_{\mathrm{rat}}$ is more strongly correlated with both $X_{\mathrm{off}}$ and $f_{\mathrm{sub}}$ for the TNG300 level 1 sample compared to the MACSIS and BAHAMAS sample.
The different redshift evolution of the correlations highlights the impact of the chosen numerical resolution and subgrid physics implementation.
The average cluster mass of all samples decreases with redshift, and due to the self-similar nature of large-scale structure formation, the substructures in each halo are less massive.
Therefore, some of these substructures are no longer resolved and this is more of an issue for lower resolution simulations such as MACSIS and BAHAMAS.
Finally, differences in the evolution of the $E_{\mathrm{rat}}$ correlations suggest that numerical choices, such as the hydrodynamics method or how AGN feedback couples energy to the ICM, are potentially impacting how quickly clusters thermalize and requires a dedicated study.
The low correlation values at $z=1.0$ between the energy ratio and the observational metrics suggest that turbulent motions within the ICM can be significant, but does not lead to significant asymmetries in the photon distribution. 

The strongest correlations in the observational metrics are between the SPA criteria.
For the BAHAMAS and TNG300 level 1 samples we find coefficients $0.75-0.9$, which reduces slightly for the MACSIS sample to $0.6-0.7$ at $z=0.1$.
This suggests there may be a slight mass dependence to the correlation, with more massive haloes yielding smaller correlation coefficients.
The correlations between the SPA criteria gradually weaken with redshift, returning coefficient values in the range $0.5-0.6$ for all samples at $z=1.0$.
Given the variation of peakiness statistic with mass, redshift, numerical resolution and subgrid physics that we found in Section \ref{sec:dist_sim}, it is interesting that it remains highly correlated with $s$ and $a$, even though they were largely insensitive to these parameters.
The SPA criteria are observational metrics that correlate least with the theoretical criteria.
At $z=0.1$, the correlation with $X_{\mathrm{off}}$, $f_{\mathrm{sub}}$ and $E_{\mathrm{rat}}$ yields values of in the range $0.3-0.4$ and, with the exception of $E_{\mathrm{rat}}$, the correlation strength remains roughly constant with redshift.

The centroid shift criterion, \cshift, is well correlated with many observable metrics, especially the photon asymmetry statistic, $A_{\mathrm{phot}}$.
The coefficient between \cshift and $A_{\mathrm{phot}}$ is $\sim0.8$ for all simulated samples at all redshifts, likely reflecting the fact that both metrics are a measure of how symmetric the emission is within a series of radial apertures.
Interestingly, the correlation between the \cshift and $s$ increases with redshift for all samples and returns coefficients $\gtrsim 0.9$ at $z=1.0$.
The centroid shift is also reasonably correlated with the theoretical criteria, yielding values of $0.6-0.8$ at $z=0.1$ for all simulated samples.
Neglecting the significant changes in $E_{\mathrm{rat}}$, the amplitude of the correlation with the theoretical metrics exhibits little redshift evolution.

The third-order power ratio, $P_{3}\,/\,P_{0}$, appears to be the least correlated metric at low redshift, producing correlation coefficients between $0.3-0.5$ for most criteria.
For the MACSIS sample, the power ratio is very weakly correlated with the theoretical criteria and the photon asymmetry parameter, with coefficients in the range $0.1-0.2$.
Given the subgrid physics and numerical resolution of the BAHAMAS and MACSIS samples is identical, the change in correlation coefficients suggests that the correlation between these metrics has some mass dependence to it.
The lack of correlation between $P_{3}\,/\,P_{0}$ and $A_{\mathrm{phot}}$ is somewhat surprising given that they are both measuring the asymmetry of the photon distribution, but our results confirm those found in a previous observational study \citep{Nurgaliev2013}.
At $z=1.0$, the power ratio becomes slightly more correlated with the other morphological metrics, with typical values in the range $0.4-0.5$.

At $z=0.1$, the photon asymmetry parameter is relatively well correlated with the theoretical metrics relative to other observational parameters, yielding correlation coefficients of $0.5-0.6$ for all samples.
These values decrease with increasing redshift, with the centre of mass offset remaining the most correlated.

In summary, we find that all $9$ metrics are positively correlated with each other at low redshift, and the correlations typically decrease with increasing redshift.
Observational parameters, except for the power ratio, are more correlated with other $2$D aperture metrics than the theoretical criteria and the opposite is true for theoretical criteria.
This is not unsurprising given that observational criteria target either strong central emission or uniformity of azimuthal mission, while theoretical criteria are focused on the presence of structure within the ICM.
However, this raises the question that, if we select relaxed cluster subsets via different morphological criteria, how consistent are they?
We conclude this work by exploring this question, which has significant implications when comparing results from different studies.

\section{Consistency}
\label{sec:consis}

\begin{figure*}
 \centering
 \includegraphics[width=\textwidth]{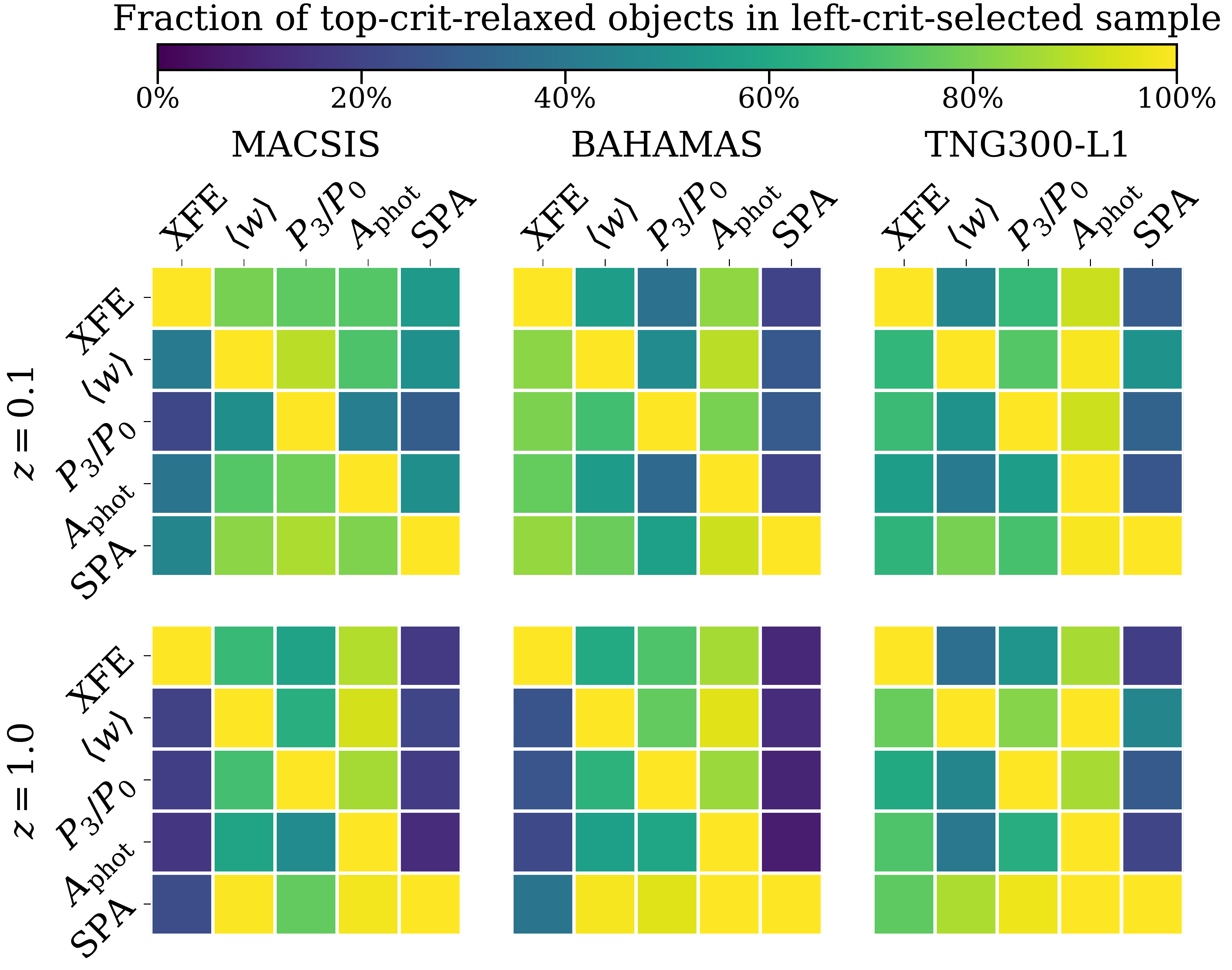}
 \caption{Consistency matrices for the $5$ relaxation criteria/combinations at $z=0.1$ (top row) and $z=1.0$ (bottom row). Results for the MACSIS (left), BAHAMAS (middle) and TNG300 level 1 (right) samples are presented. Each matrix element shows the fraction of relaxed objects according to the top criterion/combination in the subset selected by the left criterion/combination. Note, if the left criterion selects zero clusters then we denote the fraction as zero.}
 \label{fig:consis}
\end{figure*}

In this section, we assess the consistency of relaxed subsets yielded by different criteria.
Following \citet{Mantz2014}, we combine the SPA criteria into a single metric for classifying clusters as relaxed.
Additionally, we also combine the theoretical criteria into a single metric, which we label as ``XFE'', as is commonly done in many numerical studies \citep[e.g.][]{Neto2007}.
As demonstrated in Section \ref{sec:dist_sim}, many metrics show significant redshift evolution that makes a single value threshold of limited use.
For example, the energy ratio criterion defines zero clusters as relaxed in many samples at $z=1.0$.
Therefore, we now incorporate the best-fit $\beta$ values, via a $(1+z)^{-\beta}$ term, to try and ensure a given criterion defines roughly the same fraction of clusters as relaxed at a given redshift.
We note that we calibrate the peakiness metric to $z=1.0$, otherwise the given literature threshold value defines essentially no clusters as relaxed.
As previously noted, this criterion is one of the most sensitive to the subgrid implementation of AGN feedback, which remains relatively crude in the vast majority of numerical simulations.
Figure \ref{fig:consis} shows the fractions of clusters classified as relaxed by the top criterion in a subset of clusters designated as relaxed by the left criterion.
Again, we have explored the trends at all redshifts for the $5$ simulated samples, but highlight these samples and redshifts in Figure \ref{fig:consis} to demonstrate how they change with redshift, mass and subgrid physics.
Appendix \ref{app:consis} presents the complete heat maps for all simulation samples and redshifts.

If two metrics, $i$ and $j$, yield subsets composed of identical cluster samples then the fraction classified as relaxed by $j$ in the $i$th subset will be unity.
If one of the metrics is more restrictive in classifying clusters as relaxed, we will find a large fraction of the $i$th subset is defined as relaxed by $j$, but a very small fraction of the $j$th subset will be relaxed by metric $i$.
Finally, if both $i$ and $j$ are equally restrictive, but there is intrinsic scatter in the $i$-$j$ plane, we will find roughly equivalent values less than unity for both metrics.

For the subset classified as relaxed by the combined XFE criteria, we find that $80$, $75$ and $74$ per cent of the sample are also defined as relaxed by the centroid shift, power ratio and asymmetry parameters, respectively, for the MACSIS sample at $z=0.1$.
However, when examining subsets defined by these metrics we find that only $41$, $22$ and $38$ per cent of the sample are classified as relaxed by the XFE criteria.
This demonstrates that the combined XFE criteria are more restrictive and yields a smaller subset than these observable metrics.
In this case, the energy ratio criterion is driving the restrictive nature of the XFE metric.
For the BAHAMAS sample, we find the opposite is true, with the observational metrics being more selective when classifying clusters.
The energy ratio is more restrictive for the MACSIS sample because the average cluster mass is larger.
Therefore, the clusters are more recently formed relative to the BAHAMAS sample and have had less time to thermalize.
Additionally, the TNG300 level 1 sample provides a middle ground with $\sim60$ per cent of the sample defined as relaxed, regardless of whether the XFE criteria is used to select or classify the subset.
We find that the overlap between the centroid shift, power ratio and asymmetry subsets and the theoretically defined subset is dependent on the mass of the sample and the subgrid physics of the galaxy formation model.

For the subset selected by the combined SPA criteria, we find that that $\sim50-80$ per cent of the subset is classified as relaxed by the other metrics considered in this work for all of the simulated samples.
We note that for the TNG300 level 1 sample $13$ per cent of the sample is classified as relaxed by the SPA criteria at $z=0.1$.
However, when we examine the fraction classified as relaxed by SPA in subsets defined by the other metrics it yields values $<40$ per cent.
These results hold at $z=1.0$, where we find the majority of the SPA selected subset are classified as relaxed by the other metrics, but very few of the clusters in the subsets selected by these metrics are classified as relaxed by the SPA criteria.
A slight exception is the XFE combined metric subset, but then the fraction defined as relaxed depends strongly on the simulated sample, with $54$, $20$ and $29$ per cent classified as relaxed by the SPA criteria for the MACSIS, BAHAMAS and TNG300 level 1 samples, respectively, at $z=0.1$.

If we only consider the centroid shift, power ratio and asymmetry parameters, then we again find a very similar result.
Those subsets defined by one metric where the majority of the clusters are also classified as relaxed by another metric, typically yield very low relaxed percentages when the second metric is used to select the subset and the first criterion is used to classify the subset.
For all criteria, as the redshift of the MACSIS and BAHAMAS samples increases the overlap between the relaxed subsets reduces.
This is best seen for the combined XFE and SPA criteria.
On the other hand, the TNG300 level 1 sample yields consistent results at $z=0.1$ and $z=1.0$.
Given that most galaxy formation models are calibrated at $z=0.1$, the difference between samples likely highlights differences in cluster formation driven by the subgrid physics.

The lack of consistency between subsets of clusters defined as relaxed by the various morphological metrics considered in this study is partially the result of how relaxation criteria are constructed.
Though they all target some cluster feature associated with the dynamically relaxed clusters, as shown by their positive correlations, the exact threshold value of a given study is set somewhat arbitrarily and often varies between studies even for the same metric \citep[e.g.][]{Maughan2012,Weissmann2013}.
This leads to significant variations in the fraction of the cluster sample that is classified as relaxed by the different morphological metrics.
This makes defining a consistent set of relaxed clusters with current thresholds is very difficult.
Additionally, though we have shown that the relaxation criteria are correlated with each other, there is significant scatter in the plane of any two metrics.
This scatter, combined with the current thresholds, also reduces the overlap between relaxed subsets, limiting their consistency.
Therefore, the use of relaxed cluster subsets introduces significant, non-trivial selection effects that should be accounted for.
Two relaxed subsets selected by different morphological metrics may have very different properties, something that we will assess in Cao et al. (in prep.) via multivariate classification methods.

\section{Conclusions}
\label{sec:concs}
In this work, we have explored the distribution, correlation and consistency of many commonly used theoretical and observational metrics that assess the dynamical state of a galaxy cluster.
We used $5$ simulated galaxy cluster samples drawn from the BAHAMAS, MACSIS and IllustrisTNG simulations suites, enabling us to explore the impact of cluster mass and numerical choices, like hydrodynamical method and calibrated subgrid physics, on relaxation criteria.
At four redshifts ($z = 0.1$, $0.3$, $0.5$ and $1.0$) we extracted all haloes with a mass $M_{\mathrm{200,crit}}\geq10^{14}\,\mathrm{M}_{\astrosun}$.
We then generated $6$ projected synthetic X-ray images for every cluster in each sample, yielding a total of $54,096$ images, and compute $9$ commonly used morphological metrics from these images.
We then explored the distribution and correlation of the metrics as a function of mass, redshift and numerical choices for each the simulated samples and compared to observed cluster samples.
Finally, we explored the consistency of relaxed subsets defined by different criteria for the simulated samples.
Our main results are as follows:
\begin{itemize}
 \item Extracting observational data from \citet{Mantz2015}, \citet{Lovisari2017} and \citet{Nurgaliev2017}, we select and trim the MACSIS sample to match the median mass or temperature of the sample and the redshift distribution to the corresponding observational data.
 In general, we find reasonable agreement between the simulated and observed relaxation criteria distributions.
 All distributions are well described by log-normal functions.
 The simulated distribution tends to wider than the observed distribution, but we do not find major offsets between the simulated and observed distributions.
 \item We find that many criteria manifest clear evolution with redshift, which leads to significant evolution in the fraction of clusters classified as relaxed when using the fixed threshold values adopted by most studies.
 We compute the best-fit relations of the form $\propto(1+z)^{\beta}$ for each metric to quantify the average evolution across the $5$ simulation samples.
 The substructure mass fraction is the only criterion that has a negligible redshift evolution, consistent with clusters being self-similar objects.
 The criteria distributions can also be impacted by the chosen numerical resolution (e.g. substructure mass fraction) and subgrid physics (e.g. energy ratio).
 \item The correlation of the morphological metrics for the simulated samples are in reasonable agreement with observed correlations extracted from \citet{Mantz2015}, \citet{Lovisari2017} and \citet{Nurgaliev2017}.
 The trend of certain metrics, such as centroid shift and photon asymmetry, being more correlated than other, like the peakiness and alignment parameters, is reproduced by the simulations, though the simulated correlations tend to be stronger than observed.
 \item All relaxation criteria studied in this work are positively correlated with each other.
 As expected, at $z=0.1$ theoretical criteria are most strongly correlated with other theoretical criteria and the SPA criteria are strongly correlated with themselves.
 The SPA criteria are least correlated with the theoretical metrics.
 The correlation of some criteria, such as the power ratio, shows a dependence on the mass of the sample, with the MACSIS sample yielding weaker correlations relative to BAHAMAS.
 The correlations between the BAHAMAS sample and the TNG300 samples are generally in good agreement with each other.
 \item At high redshift $(z=1.0)$, we find that the correlations between the different metrics weaken relative to the low redshift $(z=0.1)$.
 However, certain combinations, like the symmetry-centroid shift and photon asymmetry-centroid shift, remain correlated to a similar degree, or even increases in strength.
 We also see differences between the simulated samples, with the energy ratio criterion being significantly less correlated with other metrics for MACSIS and BAHAMAS relative to the TNG300 samples.
 This suggests that numerical choices, such as subgrid physics implementation, is impacting the recovered correlation at high redshift for some metrics.
 \item Finally, we explored the consistency of relaxed cluster subsets produced by the $9$ relaxation criteria considered in this study.
 Removing the redshift evolution so that the same fraction of clusters are classified as relaxed at a given redshift, we explored the fraction of clusters defined as relaxed by one metric in a subset created by another.
 Due to the arbitrary nature of how threshold values are set, we find that certain criteria are far more restrictive in defining relaxed clusters and thus the consistency of relaxed cluster subsets varies significantly depending on the chosen criteria.
 These issues are further exacerbated by the intrinsic scatter present in the criterion-criterion plane of two morphological metrics.
 Therefore, the comparison of two relaxed subsets defined by different morphological metrics is non-trivial due to the selection effects introduced.
\end{itemize}
The use of morphological metrics is common to many galaxy cluster studies because it is thought to yield more precise, less biased mass estimates.
In both observational and theoretical work, there are a plethora of criteria used to define relaxed clusters.
However, the use of these metrics introduces significant selection effects due to their arbitrary threshold values, redshift evolution and intrinsic scatter.
In future work (Cao et al. in prep.), we will explore how to define consistent threshold values and explore the hyper-dimensional space composed of the different relaxation criteria to build consistent sets of relaxed galaxy clusters.

\section*{Acknowledgements}
We thank Michael McDonald and Hui Li for useful discussions and insightful comments. 
MV and DB acknowledge support through an MIT RSC award, a Kavli Research Investment Fund, NASA ATP grant NNX17AG29G, and NSF grants AST-1814053, AST-1814259 and AST-1909831.
This work used the Odyssey Cluster at Harvard University, operated by the Faculty of Arts and Sciences Research Computing (FASRC).
Additionally, this work used the DiRAC@Durham facility managed by the Institute for Computational Cosmology on behalf of the STFC DiRAC HPC Facility (www.dirac.ac.uk).
The equipment was funded by BEIS capital funding via STFC capital grants ST/K00042X/1, ST/P002293/1, ST/R002371/1 and ST/S002502/1, Durham University and STFC operations grant ST/R000832/1. DiRAC is part of the National e-Infrastructure.

\bibliographystyle{mnras}
\bibliography{ms}

\appendix

\section{Distribution parameters}
\label{app:dist}

\renewcommand\arraystretch{1.06}
\begin{table*}
 \caption{Table showing the best-fit log-normal distribution parameters for centre of mass offset ($X_\mathrm{off}$),
 using the Gaussian CDF fitting method introduced in Section \ref{sec:CDF_fit}.
 The fitting parameters ($\mu$ and $\sigma$) are shown with $1\sigma$ uncertainties and the $\chi^2$ in the chi-squared analysis.}
 \centering
 \begingroup
 \setlength{\tabcolsep}{2pt}
 \begin{tabular}{c c c c c c c c c c c c c}
  \hline
  Simulation & \multicolumn{3}{c}{$z = 0.1$} & \multicolumn{3}{c}{$z = 0.3$} & \multicolumn{3}{c}{$z = 0.5$} & \multicolumn{3}{c}{$z = 1.0$} \vspace{-1ex} \\
  & $\mu$ & $\sigma$ & $\chi^2$ & $\mu$ & $\sigma$ & $\chi^2$ & $\mu$ & $\sigma$ & $\chi^2$ & $\mu$ & $\sigma$ & $\chi^2$ \\
  \hline
  BAHAMAS & $-1.392 \!\pm\! 0.002$ & $0.355 \!\pm\! 0.002$ & $1.449$ & $-1.337 \!\pm\! 0.003$ & $0.337 \!\pm\! 0.003$ & $2.290$ & $-1.284 \!\pm\! 0.003$ & $0.321 \!\pm\! 0.003$ & $2.154$ & $-1.207 \!\pm\! 0.003$ & $0.297 \!\pm\! 0.003$ & $1.548$ \\
  MACSIS & $-1.294 \!\pm\! 0.002$ & $0.333 \!\pm\! 0.003$ & $0.984$ & $-1.271 \!\pm\! 0.004$ & $0.321 \!\pm\! 0.004$ & $1.653$ & $-1.221 \!\pm\! 0.004$ & $0.293 \!\pm\! 0.004$ & $1.618$ & $-1.186 \!\pm\! 0.003$ & $0.301 \!\pm\! 0.003$ & $1.559$ \\
  TNG300-L1 & $-1.410 \!\pm\! 0.003$ & $0.341 \!\pm\! 0.003$ & $0.913$ & $-1.314 \!\pm\! 0.004$ & $0.346 \!\pm\! 0.004$ & $1.096$ & $-1.296 \!\pm\! 0.003$ & $0.328 \!\pm\! 0.003$ & $0.887$ & $-1.201 \!\pm\! 0.008$ & $0.291 \!\pm\! 0.008$ & $1.491$ \\
  TNG300-L2 & $-1.414 \!\pm\! 0.002$ & $0.328 \!\pm\! 0.003$ & $0.760$ & $-1.350 \!\pm\! 0.004$ & $0.340 \!\pm\! 0.004$ & $1.284$ & $-1.303 \!\pm\! 0.002$ & $0.317 \!\pm\! 0.002$ & $0.667$ & $-1.218 \!\pm\! 0.010$ & $0.326 \!\pm\! 0.009$ & $1.583$ \\
  TNG300-L3 & $-1.402 \!\pm\! 0.002$ & $0.322 \!\pm\! 0.002$ & $0.535$ & $-1.303 \!\pm\! 0.003$ & $0.352 \!\pm\! 0.003$ & $0.766$ & $-1.194 \!\pm\! 0.002$ & $0.284 \!\pm\! 0.002$ & $0.571$ & $-1.042 \!\pm\! 0.005$ & $0.243 \!\pm\! 0.007$ & $1.333$ \\
  \hline
 \end{tabular}
 \endgroup
 \label{tab:Xoff_dist}
\end{table*}\begin{table*}
 \caption{Table showing the best-fit log-normal distribution parameters for substructure mass fraction ($f_\mathrm{sub}$),
 using the Gaussian CDF fitting method introduced in Section \ref{sec:CDF_fit}.
 The fitting parameters ($\mu$ and $\sigma$) are shown with $1\sigma$ uncertainties and the $\chi^2$ in the chi-squared analysis.}
 \centering
 \begingroup
 \setlength{\tabcolsep}{2pt}
 \begin{tabular}{c c c c c c c c c c c c c}
  \hline
  Simulation & \multicolumn{3}{c}{$z = 0.1$} & \multicolumn{3}{c}{$z = 0.3$} & \multicolumn{3}{c}{$z = 0.5$} & \multicolumn{3}{c}{$z = 1.0$} \vspace{-1ex} \\
  & $\mu$ & $\sigma$ & $\chi^2$ & $\mu$ & $\sigma$ & $\chi^2$ & $\mu$ & $\sigma$ & $\chi^2$ & $\mu$ & $\sigma$ & $\chi^2$ \\
  \hline
  BAHAMAS & $-1.307 \!\pm\! 0.001$ & $0.244 \!\pm\! 0.001$ & $1.059$ & $-1.283 \!\pm\! 0.001$ & $0.243 \!\pm\! 0.001$ & $1.324$ & $-1.266 \!\pm\! 0.002$ & $0.223 \!\pm\! 0.002$ & $1.894$ & $-1.265 \!\pm\! 0.002$ & $0.214 \!\pm\! 0.002$ & $1.436$ \\
  MACSIS & $-1.246 \!\pm\! 0.001$ & $0.169 \!\pm\! 0.001$ & $0.884$ & $-1.234 \!\pm\! 0.001$ & $0.174 \!\pm\! 0.001$ & $0.746$ & $-1.229 \!\pm\! 0.001$ & $0.173 \!\pm\! 0.001$ & $0.949$ & $-1.237 \!\pm\! 0.001$ & $0.185 \!\pm\! 0.001$ & $0.832$ \\
  TNG300-L1 & $-1.293 \!\pm\! 0.002$ & $0.177 \!\pm\! 0.002$ & $1.058$ & $-1.249 \!\pm\! 0.001$ & $0.176 \!\pm\! 0.001$ & $0.619$ & $-1.213 \!\pm\! 0.001$ & $0.176 \!\pm\! 0.001$ & $0.640$ & $-1.198 \!\pm\! 0.001$ & $0.141 \!\pm\! 0.002$ & $0.514$ \\
  TNG300-L2 & $-1.399 \!\pm\! 0.001$ & $0.221 \!\pm\! 0.002$ & $0.770$ & $-1.358 \!\pm\! 0.002$ & $0.203 \!\pm\! 0.002$ & $0.938$ & $-1.326 \!\pm\! 0.002$ & $0.200 \!\pm\! 0.002$ & $0.987$ & $-1.296 \!\pm\! 0.002$ & $0.171 \!\pm\! 0.003$ & $1.105$ \\
  TNG300-L3 & $-1.676 \!\pm\! 0.002$ & $0.356 \!\pm\! 0.003$ & $0.726$ & $-1.663 \!\pm\! 0.006$ & $0.302 \!\pm\! 0.004$ & $1.710$ & $-1.584 \!\pm\! 0.002$ & $0.311 \!\pm\! 0.003$ & $0.562$ & $-1.530 \!\pm\! 0.007$ & $0.304 \!\pm\! 0.007$ & $1.173$ \\
  \hline
 \end{tabular}
 \endgroup
 \label{tab:Fsub_dist}
\end{table*}\begin{table*}
 \caption{Table showing the best-fit log-normal distribution parameters for energy ratio ($E_\mathrm{rat}$),
 using the Gaussian CDF fitting method introduced in Section \ref{sec:CDF_fit}.
 The fitting parameters ($\mu$ and $\sigma$) are shown with $1\sigma$ uncertainties and the $\chi^2$ in the chi-squared analysis.}
 \centering
 \begingroup
 \setlength{\tabcolsep}{2pt}
 \begin{tabular}{c c c c c c c c c c c c c}
  \hline
  Simulation & \multicolumn{3}{c}{$z = 0.1$} & \multicolumn{3}{c}{$z = 0.3$} & \multicolumn{3}{c}{$z = 0.5$} & \multicolumn{3}{c}{$z = 1.0$} \vspace{-1ex} \\
  & $\mu$ & $\sigma$ & $\chi^2$ & $\mu$ & $\sigma$ & $\chi^2$ & $\mu$ & $\sigma$ & $\chi^2$ & $\mu$ & $\sigma$ & $\chi^2$ \\
  \hline
  BAHAMAS & $-0.982 \!\pm\! 0.001$ & $0.189 \!\pm\! 0.001$ & $1.225$ & $-0.885 \!\pm\! 0.000$ & $0.176 \!\pm\! 0.001$ & $0.792$ & $-0.724 \!\pm\! 0.000$ & $0.161 \!\pm\! 0.001$ & $0.750$ & $-0.473 \!\pm\! 0.001$ & $0.193 \!\pm\! 0.001$ & $0.708$ \\
  MACSIS & $-0.745 \!\pm\! 0.001$ & $0.208 \!\pm\! 0.001$ & $0.857$ & $-0.705 \!\pm\! 0.001$ & $0.180 \!\pm\! 0.001$ & $0.971$ & $-0.669 \!\pm\! 0.001$ & $0.171 \!\pm\! 0.001$ & $0.684$ & $-0.438 \!\pm\! 0.001$ & $0.194 \!\pm\! 0.001$ & $0.827$ \\
  TNG300-L1 & $-0.889 \!\pm\! 0.002$ & $0.234 \!\pm\! 0.002$ & $0.944$ & $-0.835 \!\pm\! 0.002$ & $0.213 \!\pm\! 0.002$ & $0.840$ & $-0.774 \!\pm\! 0.001$ & $0.215 \!\pm\! 0.001$ & $0.433$ & $-0.661 \!\pm\! 0.002$ & $0.152 \!\pm\! 0.003$ & $0.860$ \\
  TNG300-L2 & $-0.874 \!\pm\! 0.003$ & $0.216 \!\pm\! 0.003$ & $1.255$ & $-0.830 \!\pm\! 0.001$ & $0.212 \!\pm\! 0.002$ & $0.692$ & $-0.762 \!\pm\! 0.001$ & $0.211 \!\pm\! 0.001$ & $0.484$ & $-0.643 \!\pm\! 0.002$ & $0.153 \!\pm\! 0.002$ & $0.683$ \\
  TNG300-L3 & $-0.879 \!\pm\! 0.002$ & $0.203 \!\pm\! 0.002$ & $0.951$ & $-0.821 \!\pm\! 0.001$ & $0.199 \!\pm\! 0.001$ & $0.620$ & $-0.766 \!\pm\! 0.001$ & $0.193 \!\pm\! 0.001$ & $0.365$ & $-0.659 \!\pm\! 0.003$ & $0.165 \!\pm\! 0.003$ & $0.974$ \\
  \hline
 \end{tabular}
 \endgroup
 \label{tab:Erat_dist}
\end{table*}\begin{table*}
 \caption{Table showing the best-fit log-normal distribution parameters for centroid shift ($\langle w \rangle$),
 using the Gaussian CDF fitting method introduced in Section \ref{sec:CDF_fit}.
 The fitting parameters ($\mu$ and $\sigma$) are shown with $1\sigma$ uncertainties and the $\chi^2$ in the chi-squared analysis.}
 \centering
 \begingroup
 \setlength{\tabcolsep}{2pt}
 \begin{tabular}{c c c c c c c c c c c c c}
  \hline
  Simulation & \multicolumn{3}{c}{$z = 0.1$} & \multicolumn{3}{c}{$z = 0.3$} & \multicolumn{3}{c}{$z = 0.5$} & \multicolumn{3}{c}{$z = 1.0$} \vspace{-1ex} \\
  & $\mu$ & $\sigma$ & $\chi^2$ & $\mu$ & $\sigma$ & $\chi^2$ & $\mu$ & $\sigma$ & $\chi^2$ & $\mu$ & $\sigma$ & $\chi^2$ \\
  \hline
  BAHAMAS & $-2.016 \!\pm\! 0.001$ & $0.457 \!\pm\! 0.001$ & $1.881$ & $-1.969 \!\pm\! 0.002$ & $0.449 \!\pm\! 0.003$ & $3.563$ & $-1.882 \!\pm\! 0.002$ & $0.424 \!\pm\! 0.002$ & $2.930$ & $-1.761 \!\pm\! 0.003$ & $0.399 \!\pm\! 0.003$ & $2.360$ \\
  MACSIS & $-2.012 \!\pm\! 0.007$ & $0.542 \!\pm\! 0.007$ & $4.143$ & $-1.929 \!\pm\! 0.008$ & $0.503 \!\pm\! 0.008$ & $5.389$ & $-1.874 \!\pm\! 0.006$ & $0.486 \!\pm\! 0.006$ & $3.788$ & $-1.754 \!\pm\! 0.003$ & $0.426 \!\pm\! 0.004$ & $2.653$ \\
  TNG300-L1 & $-1.902 \!\pm\! 0.002$ & $0.455 \!\pm\! 0.002$ & $1.156$ & $-1.781 \!\pm\! 0.006$ & $0.429 \!\pm\! 0.006$ & $3.094$ & $-1.742 \!\pm\! 0.002$ & $0.412 \!\pm\! 0.003$ & $1.130$ & $-1.630 \!\pm\! 0.006$ & $0.361 \!\pm\! 0.005$ & $2.120$ \\
  TNG300-L2 & $-1.948 \!\pm\! 0.002$ & $0.461 \!\pm\! 0.002$ & $1.102$ & $-1.817 \!\pm\! 0.003$ & $0.431 \!\pm\! 0.004$ & $1.936$ & $-1.750 \!\pm\! 0.002$ & $0.393 \!\pm\! 0.003$ & $1.219$ & $-1.610 \!\pm\! 0.005$ & $0.328 \!\pm\! 0.005$ & $2.019$ \\
  TNG300-L3 & $-1.978 \!\pm\! 0.002$ & $0.463 \!\pm\! 0.003$ & $1.250$ & $-1.878 \!\pm\! 0.002$ & $0.471 \!\pm\! 0.003$ & $1.356$ & $-1.743 \!\pm\! 0.002$ & $0.386 \!\pm\! 0.002$ & $1.139$ & $-1.614 \!\pm\! 0.008$ & $0.346 \!\pm\! 0.008$ & $2.452$ \\
  \hline
 \end{tabular}
 \endgroup
 \label{tab:Wshift_dist}
\end{table*}\begin{table*}
 \caption{Table showing the best-fit log-normal distribution parameters for power ratio ($P_{3}/P_{0}$),
 using the Gaussian CDF fitting method introduced in Section \ref{sec:CDF_fit}.
 The fitting parameters ($\mu$ and $\sigma$) are shown with $1\sigma$ uncertainties and the $\chi^2$ in the chi-squared analysis.}
 \centering
 \begingroup
 \setlength{\tabcolsep}{2pt}
 \begin{tabular}{c c c c c c c c c c c c c}
  \hline
  Simulation & \multicolumn{3}{c}{$z = 0.1$} & \multicolumn{3}{c}{$z = 0.3$} & \multicolumn{3}{c}{$z = 0.5$} & \multicolumn{3}{c}{$z = 1.0$} \vspace{-1ex} \\
  & $\mu$ & $\sigma$ & $\chi^2$ & $\mu$ & $\sigma$ & $\chi^2$ & $\mu$ & $\sigma$ & $\chi^2$ & $\mu$ & $\sigma$ & $\chi^2$ \\
  \hline
  BAHAMAS & $-7.528 \!\pm\! 0.006$ & $0.437 \!\pm\! 0.010$ & $8.990$ & $-7.118 \!\pm\! 0.005$ & $0.413 \!\pm\! 0.008$ & $7.581$ & $-6.857 \!\pm\! 0.003$ & $0.364 \!\pm\! 0.005$ & $4.964$ & $-6.715 \!\pm\! 0.004$ & $0.390 \!\pm\! 0.006$ & $3.739$ \\
  MACSIS & $-8.395 \!\pm\! 0.018$ & $1.044 \!\pm\! 0.019$ & $5.135$ & $-6.902 \!\pm\! 0.009$ & $1.181 \!\pm\! 0.011$ & $2.643$ & $-6.458 \!\pm\! 0.007$ & $1.052 \!\pm\! 0.010$ & $2.310$ & $-6.552 \!\pm\! 0.005$ & $0.497 \!\pm\! 0.007$ & $3.376$ \\
  TNG300-L1 & $-7.690 \!\pm\! 0.007$ & $0.447 \!\pm\! 0.009$ & $3.454$ & $-6.976 \!\pm\! 0.006$ & $0.420 \!\pm\! 0.008$ & $3.421$ & $-6.803 \!\pm\! 0.007$ & $0.310 \!\pm\! 0.008$ & $3.994$ & $-6.706 \!\pm\! 0.007$ & $0.325 \!\pm\! 0.007$ & $2.560$ \\
  TNG300-L2 & $-7.742 \!\pm\! 0.007$ & $0.461 \!\pm\! 0.010$ & $3.820$ & $-7.047 \!\pm\! 0.006$ & $0.412 \!\pm\! 0.007$ & $3.162$ & $-6.794 \!\pm\! 0.007$ & $0.345 \!\pm\! 0.009$ & $3.729$ & $-6.685 \!\pm\! 0.008$ & $0.332 \!\pm\! 0.008$ & $2.747$ \\
  TNG300-L3 & $-7.913 \!\pm\! 0.007$ & $0.634 \!\pm\! 0.011$ & $2.922$ & $-7.238 \!\pm\! 0.004$ & $0.544 \!\pm\! 0.005$ & $1.734$ & $-6.978 \!\pm\! 0.005$ & $0.561 \!\pm\! 0.007$ & $2.092$ & $-6.790 \!\pm\! 0.005$ & $0.565 \!\pm\! 0.008$ & $1.226$ \\
  \hline
 \end{tabular}
 \endgroup
 \label{tab:PowerRatio_dist}
\end{table*}\begin{table*}
 \caption{Table showing the best-fit log-normal distribution parameters for photon asymmetry ($A_\mathrm{phot}$),
 using the Gaussian CDF fitting method introduced in Section \ref{sec:CDF_fit}.
 The fitting parameters ($\mu$ and $\sigma$) are shown with $1\sigma$ uncertainties and the $\chi^2$ in the chi-squared analysis.}
 \centering
 \begingroup
 \setlength{\tabcolsep}{2pt}
 \begin{tabular}{c c c c c c c c c c c c c}
  \hline
  Simulation & \multicolumn{3}{c}{$z = 0.1$} & \multicolumn{3}{c}{$z = 0.3$} & \multicolumn{3}{c}{$z = 0.5$} & \multicolumn{3}{c}{$z = 1.0$} \vspace{-1ex} \\
  & $\mu$ & $\sigma$ & $\chi^2$ & $\mu$ & $\sigma$ & $\chi^2$ & $\mu$ & $\sigma$ & $\chi^2$ & $\mu$ & $\sigma$ & $\chi^2$ \\
  \hline
  BAHAMAS & $-0.871 \!\pm\! 0.003$ & $0.449 \!\pm\! 0.004$ & $4.825$ & $-0.808 \!\pm\! 0.002$ & $0.467 \!\pm\! 0.003$ & $3.670$ & $-0.754 \!\pm\! 0.002$ & $0.465 \!\pm\! 0.002$ & $2.361$ & $-0.657 \!\pm\! 0.003$ & $0.476 \!\pm\! 0.004$ & $2.241$ \\
  MACSIS & $-0.604 \!\pm\! 0.003$ & $0.465 \!\pm\! 0.003$ & $1.900$ & $-0.571 \!\pm\! 0.004$ & $0.457 \!\pm\! 0.004$ & $3.020$ & $-0.556 \!\pm\! 0.002$ & $0.438 \!\pm\! 0.003$ & $1.831$ & $-0.569 \!\pm\! 0.002$ & $0.453 \!\pm\! 0.003$ & $1.749$ \\
  TNG300-L1 & $-0.985 \!\pm\! 0.005$ & $0.539 \!\pm\! 0.004$ & $2.833$ & $-0.832 \!\pm\! 0.005$ & $0.560 \!\pm\! 0.005$ & $2.087$ & $-0.785 \!\pm\! 0.002$ & $0.543 \!\pm\! 0.003$ & $1.014$ & $-0.673 \!\pm\! 0.012$ & $0.494 \!\pm\! 0.009$ & $2.839$ \\
  TNG300-L2 & $-1.019 \!\pm\! 0.005$ & $0.500 \!\pm\! 0.006$ & $2.827$ & $-0.848 \!\pm\! 0.003$ & $0.533 \!\pm\! 0.003$ & $1.360$ & $-0.787 \!\pm\! 0.003$ & $0.517 \!\pm\! 0.004$ & $1.326$ & $-0.623 \!\pm\! 0.008$ & $0.551 \!\pm\! 0.008$ & $1.767$ \\
  TNG300-L3 & $-0.969 \!\pm\! 0.003$ & $0.505 \!\pm\! 0.004$ & $1.840$ & $-0.823 \!\pm\! 0.002$ & $0.563 \!\pm\! 0.002$ & $0.804$ & $-0.651 \!\pm\! 0.002$ & $0.507 \!\pm\! 0.003$ & $0.902$ & $-0.463 \!\pm\! 0.007$ & $0.429 \!\pm\! 0.009$ & $1.839$ \\
  \hline
 \end{tabular}
 \endgroup
 \label{tab:Aphot_dist}
\end{table*}\begin{table*}
 \caption{Table showing the best-fit normal distribution parameters for surface brightness peakiness ($p$),
 using the Gaussian CDF fitting method introduced in Section \ref{sec:CDF_fit}.
 The fitting parameters ($\mu$ and $\sigma$) are shown with $1\sigma$ uncertainties and the $\chi^2$ in the chi-squared analysis.}
 \centering
 \begingroup
 \setlength{\tabcolsep}{2pt}
 \begin{tabular}{c c c c c c c c c c c c c}
  \hline
  Simulation & \multicolumn{3}{c}{$z = 0.1$} & \multicolumn{3}{c}{$z = 0.3$} & \multicolumn{3}{c}{$z = 0.5$} & \multicolumn{3}{c}{$z = 1.0$} \vspace{-1ex} \\
  & $\mu$ & $\sigma$ & $\chi^2$ & $\mu$ & $\sigma$ & $\chi^2$ & $\mu$ & $\sigma$ & $\chi^2$ & $\mu$ & $\sigma$ & $\chi^2$ \\
  \hline
  BAHAMAS & $-1.687 \!\pm\! 0.005$ & $0.424 \!\pm\! 0.006$ & $8.781$ & $-1.476 \!\pm\! 0.004$ & $0.461 \!\pm\! 0.004$ & $6.070$ & $-1.089 \!\pm\! 0.008$ & $0.368 \!\pm\! 0.009$ & $13.142$ & $-0.687 \!\pm\! 0.002$ & $0.204 \!\pm\! 0.003$ & $4.187$ \\
  MACSIS & $-1.050 \!\pm\! 0.018$ & $0.432 \!\pm\! 0.015$ & $12.536$ & $-0.903 \!\pm\! 0.009$ & $0.407 \!\pm\! 0.009$ & $7.895$ & $-0.806 \!\pm\! 0.005$ & $0.345 \!\pm\! 0.008$ & $5.859$ & $-0.611 \!\pm\! 0.001$ & $0.187 \!\pm\! 0.002$ & $2.421$ \\
  TNG300-L1 & $-1.729 \!\pm\! 0.004$ & $0.318 \!\pm\! 0.006$ & $4.369$ & $-1.437 \!\pm\! 0.002$ & $0.279 \!\pm\! 0.003$ & $2.347$ & $-1.113 \!\pm\! 0.003$ & $0.224 \!\pm\! 0.004$ & $3.393$ & $-0.749 \!\pm\! 0.001$ & $0.096 \!\pm\! 0.001$ & $1.307$ \\
  TNG300-L2 & $-1.644 \!\pm\! 0.004$ & $0.356 \!\pm\! 0.004$ & $3.256$ & $-1.390 \!\pm\! 0.002$ & $0.311 \!\pm\! 0.002$ & $1.777$ & $-1.148 \!\pm\! 0.003$ & $0.266 \!\pm\! 0.004$ & $3.079$ & $-0.748 \!\pm\! 0.001$ & $0.108 \!\pm\! 0.001$ & $1.423$ \\
  TNG300-L3 & $-1.524 \!\pm\! 0.005$ & $0.493 \!\pm\! 0.006$ & $3.709$ & $-1.259 \!\pm\! 0.005$ & $0.395 \!\pm\! 0.004$ & $3.076$ & $-1.008 \!\pm\! 0.005$ & $0.331 \!\pm\! 0.006$ & $3.444$ & $-0.606 \!\pm\! 0.002$ & $0.169 \!\pm\! 0.003$ & $1.984$ \\
  \hline
 \end{tabular}
 \endgroup
 \label{tab:Peakiness_dist}
\end{table*}\begin{table*}
 \caption{Table showing the best-fit normal distribution parameters for symmetry statistic ($s$),
 using the Gaussian CDF fitting method introduced in Section \ref{sec:CDF_fit}.
 The fitting parameters ($\mu$ and $\sigma$) are shown with $1\sigma$ uncertainties and the $\chi^2$ in the chi-squared analysis.}
 \centering
 \begingroup
 \setlength{\tabcolsep}{2pt}
 \begin{tabular}{c c c c c c c c c c c c c}
  \hline
  Simulation & \multicolumn{3}{c}{$z = 0.1$} & \multicolumn{3}{c}{$z = 0.3$} & \multicolumn{3}{c}{$z = 0.5$} & \multicolumn{3}{c}{$z = 1.0$} \vspace{-1ex} \\
  & $\mu$ & $\sigma$ & $\chi^2$ & $\mu$ & $\sigma$ & $\chi^2$ & $\mu$ & $\sigma$ & $\chi^2$ & $\mu$ & $\sigma$ & $\chi^2$ \\
  \hline
  BAHAMAS & $0.556 \!\pm\! 0.003$ & $0.437 \!\pm\! 0.004$ & $5.041$ & $0.583 \!\pm\! 0.003$ & $0.413 \!\pm\! 0.005$ & $5.660$ & $0.671 \!\pm\! 0.003$ & $0.369 \!\pm\! 0.004$ & $4.788$ & $0.727 \!\pm\! 0.001$ & $0.317 \!\pm\! 0.001$ & $1.164$ \\
  MACSIS & $0.758 \!\pm\! 0.004$ & $0.387 \!\pm\! 0.005$ & $2.930$ & $0.714 \!\pm\! 0.002$ & $0.373 \!\pm\! 0.003$ & $1.966$ & $0.729 \!\pm\! 0.003$ & $0.351 \!\pm\! 0.003$ & $2.341$ & $0.772 \!\pm\! 0.001$ & $0.328 \!\pm\! 0.002$ & $1.271$ \\
  TNG300-L1 & $0.512 \!\pm\! 0.005$ & $0.497 \!\pm\! 0.008$ & $2.775$ & $0.591 \!\pm\! 0.003$ & $0.420 \!\pm\! 0.003$ & $1.642$ & $0.715 \!\pm\! 0.002$ & $0.348 \!\pm\! 0.003$ & $1.284$ & $0.840 \!\pm\! 0.004$ & $0.293 \!\pm\! 0.004$ & $2.016$ \\
  TNG300-L2 & $0.629 \!\pm\! 0.004$ & $0.468 \!\pm\! 0.006$ & $2.512$ & $0.629 \!\pm\! 0.004$ & $0.476 \!\pm\! 0.005$ & $2.102$ & $0.696 \!\pm\! 0.002$ & $0.376 \!\pm\! 0.003$ & $1.312$ & $0.810 \!\pm\! 0.003$ & $0.304 \!\pm\! 0.003$ & $1.111$ \\
  TNG300-L3 & $0.739 \!\pm\! 0.002$ & $0.466 \!\pm\! 0.003$ & $1.358$ & $0.737 \!\pm\! 0.005$ & $0.457 \!\pm\! 0.006$ & $2.652$ & $0.727 \!\pm\! 0.003$ & $0.363 \!\pm\! 0.003$ & $1.474$ & $0.742 \!\pm\! 0.005$ & $0.256 \!\pm\! 0.007$ & $2.323$ \\
  \hline
 \end{tabular}
 \endgroup
 \label{tab:Symmetry_dist}
\end{table*}\begin{table*}
 \caption{Table showing the best-fit normal distribution parameters for alignment statistic ($a$),
 using the Gaussian CDF fitting method introduced in Section \ref{sec:CDF_fit}.
 The fitting parameters ($\mu$ and $\sigma$) are shown with $1\sigma$ uncertainties and the $\chi^2$ in the chi-squared analysis.}
 \centering
 \begingroup
 \setlength{\tabcolsep}{2pt}
 \begin{tabular}{c c c c c c c c c c c c c}
  \hline
  Simulation & \multicolumn{3}{c}{$z = 0.1$} & \multicolumn{3}{c}{$z = 0.3$} & \multicolumn{3}{c}{$z = 0.5$} & \multicolumn{3}{c}{$z = 1.0$} \vspace{-1ex} \\
  & $\mu$ & $\sigma$ & $\chi^2$ & $\mu$ & $\sigma$ & $\chi^2$ & $\mu$ & $\sigma$ & $\chi^2$ & $\mu$ & $\sigma$ & $\chi^2$ \\
  \hline
  BAHAMAS & $0.809 \!\pm\! 0.003$ & $0.363 \!\pm\! 0.004$ & $5.577$ & $0.817 \!\pm\! 0.003$ & $0.364 \!\pm\! 0.004$ & $5.808$ & $0.859 \!\pm\! 0.004$ & $0.354 \!\pm\! 0.005$ & $6.140$ & $1.017 \!\pm\! 0.003$ & $0.322 \!\pm\! 0.003$ & $2.942$ \\
  MACSIS & $0.854 \!\pm\! 0.003$ & $0.373 \!\pm\! 0.004$ & $2.539$ & $0.855 \!\pm\! 0.002$ & $0.373 \!\pm\! 0.003$ & $2.047$ & $0.896 \!\pm\! 0.003$ & $0.369 \!\pm\! 0.003$ & $2.423$ & $1.094 \!\pm\! 0.003$ & $0.311 \!\pm\! 0.003$ & $2.722$ \\
  TNG300-L1 & $0.676 \!\pm\! 0.007$ & $0.494 \!\pm\! 0.007$ & $3.817$ & $0.792 \!\pm\! 0.005$ & $0.455 \!\pm\! 0.005$ & $2.829$ & $0.981 \!\pm\! 0.003$ & $0.394 \!\pm\! 0.004$ & $1.878$ & $1.274 \!\pm\! 0.004$ & $0.254 \!\pm\! 0.005$ & $1.838$ \\
  TNG300-L2 & $0.801 \!\pm\! 0.005$ & $0.457 \!\pm\! 0.005$ & $2.912$ & $0.865 \!\pm\! 0.005$ & $0.461 \!\pm\! 0.005$ & $2.422$ & $0.928 \!\pm\! 0.002$ & $0.413 \!\pm\! 0.003$ & $1.243$ & $1.235 \!\pm\! 0.003$ & $0.257 \!\pm\! 0.004$ & $1.396$ \\
  TNG300-L3 & $0.924 \!\pm\! 0.005$ & $0.467 \!\pm\! 0.005$ & $3.006$ & $0.939 \!\pm\! 0.006$ & $0.458 \!\pm\! 0.006$ & $3.249$ & $0.976 \!\pm\! 0.004$ & $0.408 \!\pm\! 0.005$ & $2.264$ & $1.133 \!\pm\! 0.002$ & $0.275 \!\pm\! 0.003$ & $0.900$ \\
  \hline
 \end{tabular}
 \endgroup
 \label{tab:Alignment_dist}
\end{table*}
\renewcommand\arraystretch{1.0}

Tables \ref{tab:Xoff_dist} to \ref{tab:Alignment_dist} show the best fit (log-)normal distribution parameters ($\mu$ and $\sigma$) of each relaxation parameter in each simulation at each redshift studied in this work, obtained via the Gaussian CDF fitting method introduced in Section \ref{sec:CDF_fit}, together with 1$\sigma$ uncertainties and the $\chi^2$ in the chi-squared analysis.
See Section \ref{sec:dist_sim} for further explanations.

\section{Correlation heatmaps}
\label{app:corr}

\begin{figure*}
 \includegraphics[width=0.94\textwidth]{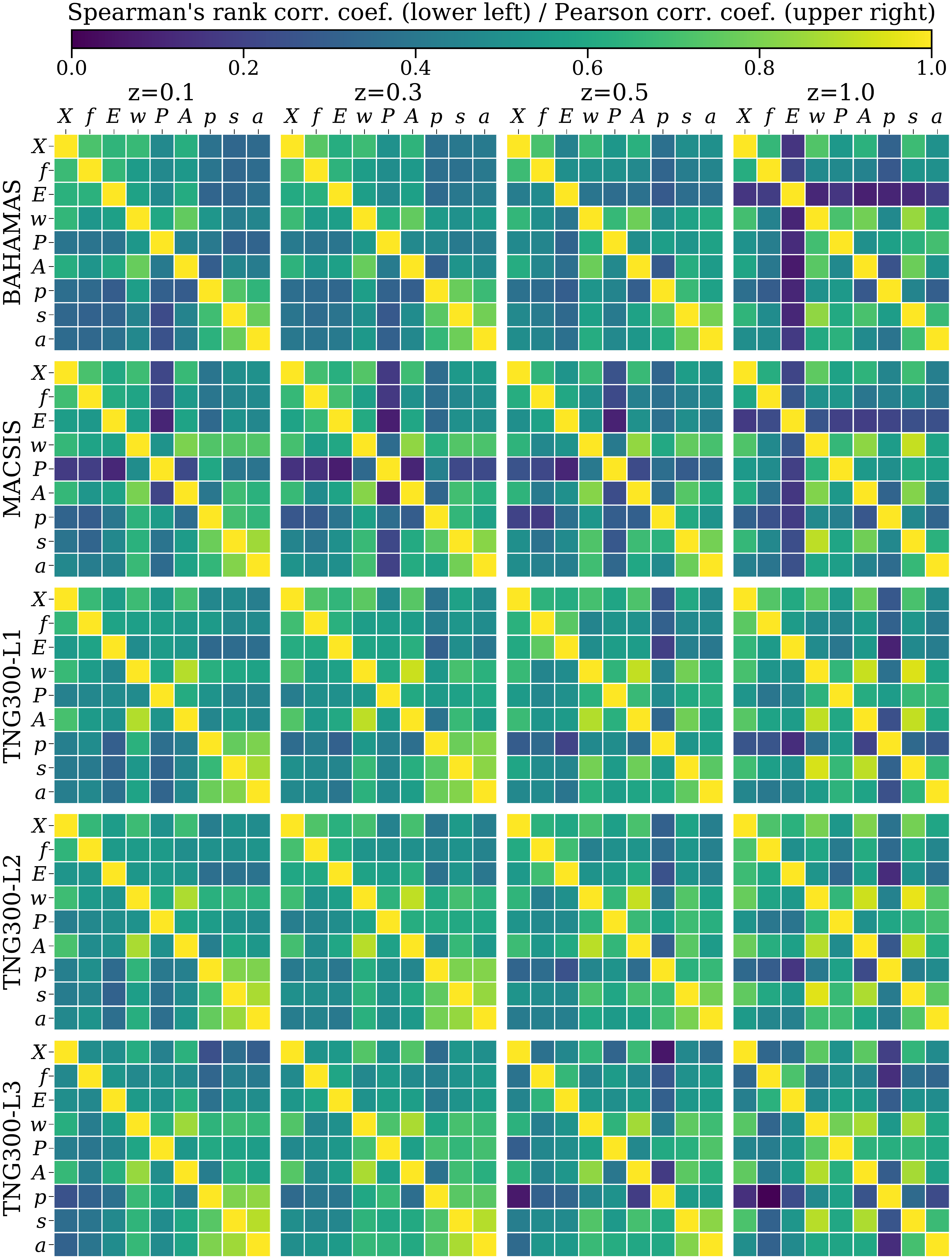}
 \caption{Extended version of Figure \ref{fig:corr} in Section \ref{sec:corr_sim}. See the caption of Figure \ref{fig:corr} for explanations.
 For simplicity, each relaxation parameter is denoted here with the principal letter in its symbol, namely:
 $X$ for centre of mass offset ($X_{\mathrm{off}}$), $f$ for substructure mass fraction ($f_{\mathrm{sub}}$), $E$ for energy ratio ($E_{\mathrm{rat}}$), $w$ for centroid shift (\cshift), $P$ for third-order power ratio ($P_{3}/P_{0}$), $A$ for Photon asymmetry ($A_{\mathrm{phot}}$)
 and $p$ for surface brightness peakiness, $s$ for symmetry statistic, $a$ for alignment statistic, as usual.}
 \label{fig:corr_ext}
\end{figure*}

Figure \ref{fig:corr_ext} is an extended version of Figure \ref{fig:corr} in Section \ref{sec:corr_sim}.
See the caption of Figure \ref{fig:corr} and the text in Section \ref{sec:corr_sim} for further explanations.

\section{Consistency heatmaps}
\label{app:consis}

\begin{figure*}
 \includegraphics[width=\textwidth]{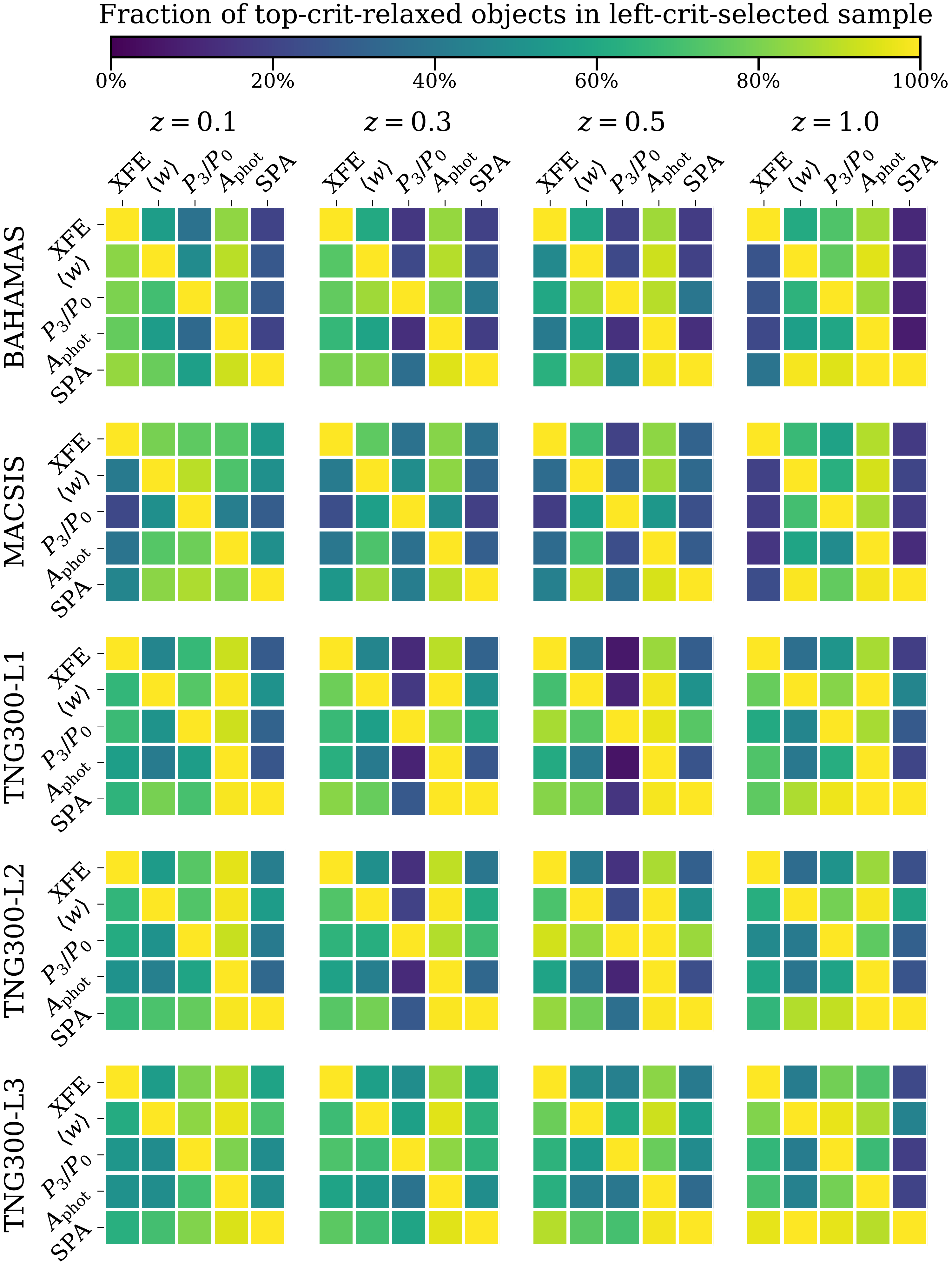}
 \caption{Extended version of Figure \ref{fig:consis} in Section \ref{sec:consis}. See the caption of Figure \ref{fig:consis} for explanations.}
 \label{fig:consis_ext}
\end{figure*}

Figure \ref{fig:consis_ext} is an extended version of Figure \ref{fig:consis} in Section \ref{sec:consis}.
See the caption of Figure \ref{fig:consis} and the text in Section \ref{sec:consis} for further explanations.

\bsp
\label{lastpage}
\end{document}